\documentclass[journal]{IEEEtran}
\usepackage{amsthm}
\usepackage{cite}
\usepackage{amsmath}
\usepackage{nccmath}
\usepackage{amssymb}
\usepackage{amsfonts}
\usepackage{graphicx}
\usepackage{fancyhdr}
\usepackage{svg}
\usepackage{bbm}
\usepackage{subcaption}
\usepackage{textcomp}
\usepackage{xcolor}
\usepackage[linesnumbered,ruled]{algorithm2e}
\usepackage{booktabs}
\usepackage{bbold}
\usepackage{kotex}

\theoremstyle{definition}

\usepackage{multicol}
\usepackage{multirow}

\usepackage[normalem]{ulem}
\usepackage{lettrine}

\begin{document}

\title{Situation-Aware Deep Reinforcement Learning for Autonomous Nonlinear Mobility Control in Cyber-Physical Loitering Munition Systems}

\author{
    Hyunsoo Lee,
    Soohyun Park, 
    Won Joon Yun, 
    Soyi Jung,~\IEEEmembership{Member, IEEE},
    and \\
    Joongheon Kim,~\IEEEmembership{Senior Member, IEEE}
    \thanks{The earlier version of this work was presented at IEEE Vehicular Technology Society (VTS) Asia Pacific Wireless Communications Symposium (APWCS), Seoul, Korea, August 2022~\cite{APWCS22lee}, and this paper received IEEE VTS Seoul Chapter Award (December 20222).} 
    \thanks{This research was funded by the National Research Foundation of Korea (2022R1A2C2004869). \textit{(Corresponding authors: Soyi Jung, Joongheon Kim)}}
    \thanks{Hyunsoo Lee, Soohyun Park, Won Joon Yun, and Joongheon Kim are with the Department of Electrical and Computer Engineering, Korea University, Seoul 02841, Republic of Korea (e-mails: \{hyunsoo, soohyun828, ywjoon95, joongheon\}@korea.ac.kr).}
    \thanks{Soyi Jung is with the Department of Electrical of Computer Engineering, Ajou University, Suwon, Republic of Korea (e-mail: sjung@ajou.ac.kr).}
}
\maketitle

\begin{abstract}
According to the rapid development of drone technologies, drones are widely used in many applications including military domains. 
In this paper, a novel situation-aware DRL-based autonomous nonlinear drone mobility control algorithm in cyber-physical loitering munition applications.
On the battlefield, the design of DRL-based autonomous control algorithm is not straightforward because real-world data gathering is generally not available. Therefore, the approach in this paper is that cyber-physical virtual environment is constructed with Unity environment. Based on the virtual cyber-physical battlefield scenarios, a DRL-based automated nonlinear drone mobility control algorithm can be designed, evaluated, and visualized. 
Moreover, many obstacles exist which is harmful for linear trajectory control in real-world battlefield scenarios. Thus, our proposed autonomous nonlinear drone mobility control algorithm utilizes situation-aware components those are implemented with a Raycast function in Unity virtual scenarios. Based on the gathered situation-aware information, the drone can autonomously and nonlinearly adjust its trajectory during flight. Therefore, this approach is obviously beneficial for avoiding obstacles in obstacle-deployed battlefields. Our visualization-based performance evaluation shows that the proposed algorithm is superior from the other linear mobility control algorithms. 

\end{abstract}

\IEEEpeerreviewmaketitle



%

\section{Introduction}\label{sec:introduction}
\subsection{Background and Motivation}

{With the Internet of Things (IoT) revolution in modern communications and network applications, drones for autonomous aerial ad-hoc on-demand three-dimensional (3D) networking have encountered a rapid change in their applications, from professional toys to professional application-specific complex IoT devices~\cite{isj1,isj2,isj3}. In modern embedded drone design and implementation, drones are integrated with several sensors, including cameras and global positioning system (GPS), and thus, they are actively and widely used in various fields. It means the drones are not only used for live broadcasts, agriculture, and weather forecasting but also have potentials to be used in global logistics or future mobility to bring dramatic convenience to human life. For more details, when drones are utilized in agriculture applications, multi-drone networking platform can manage vast farms, diagnose crop conditions, and provide appropriate solutions to increase productivity~\cite{agriculture}. In the field of logistics, enterprises like Amazon have launched drone delivery pilot services, delivering packages safely and on time through fully autonomous driving systems. Kong \textit{et al.}~\cite{logistics} also conducted a research project to optimize the drone path/trajectory using attention-based pointer networks for future mobility applications. In future mobility, when vertiports are built in smart cities, drone taxis are expected to transport passengers~\cite{icte21yun}. 


\begin{figure}[t!]
    \centering
    \includegraphics[width =0.99\linewidth]{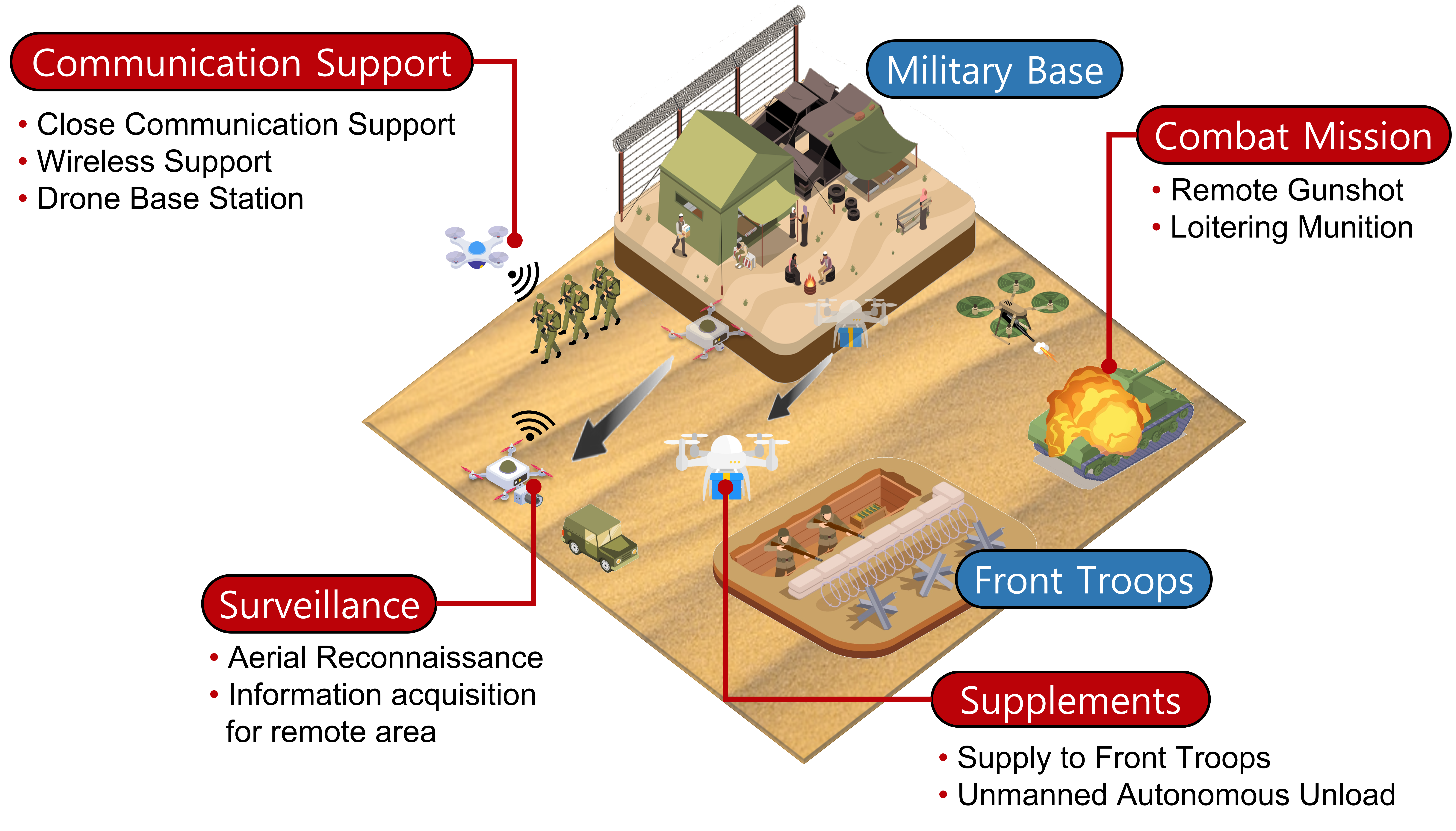}
    \caption{The use of drones in the military battlefield, i.e., autonomous drone mobility control for loitering munition.}
    	\label{fig:military}
\end{figure}
In order to fully utilize the drone-related technologies in many emerging applications, research to control the drone mobility in an autonomous way is essentially required. 
To design and implement the algorithms for the autonomous drone mobility control, the use of \textit{deep reinforcement learning (DRL)} is one of the promising deep learning algorithms because DRL is formally defined as a discrete-time stochastic decision-making control process for maximizing the expected return/utility. 
Furthermore, during drone trajectory control in various applications, many obstacles (such as buildings and structures) can exist. Therefore, nonlinear autonomous drone mobility control under the consideration of near-field \textit{situation sensing} is essentially desired for avoiding the obstacles in physical worlds.
Although major studies have been conducted on how drones avoid obstacles via situation-aware DRL algorithms, this research contribution is inappropriate for military usage (such as loitering munition) as it has been studied in terms of multi-drone swarm flight for attacking target areas via autonomous nonlinear drone mobility control~\cite{obstacle, obstacle2}.



In considering multi-drone autonomous networks, the use of cyber-physical systems for interacting/mapping between the actions in the virtual environment and the ones in the real world environment is getting a lot of attention in industry and academia~\cite{survey_icps, access_cps, tran2019safety, CysWater}. 
The CPS technologies are based on the integration of communications, networking, sensing, mechatronics, and data analytics. Therefore, CPS is widely used in many industrial applications such as smart production, smart electric grids, smart logistics, and smart health care~\cite{CysWater, CysWater[1]}.
In order to realize the interaction/mapping between the actions in the virtual environment and the ones in the physical environment, visualization tools (such as Unity) can be definitely helpful because the algorithm designers can intuitively and immediately understand and validate the algorithm functionalities, as well-studied in~\cite{tran2019safety, CPS-unity1, CPS-unity2, CPS-unity3}.

 \subsection{Use Cases}
In this paper, among various promising applications and use cases in order to utilize DRL algorithms in multi-drone networks, military applications are considered because it is one of the promising topics nowadays based on the big success in modern wars, especially in urban areas. 
In the military field, the application-specific drones can take on a variety of roles, such as monitoring the enemy, providing supplies, and executing combat missions, as illustrated in Fig.~\ref{fig:military}. Despite the short operation time due to the limitations of batteries (approximately a few tens of minutes), military drones have a significant advantage in swiftly navigating areas that infantry cannot observe, with avoiding pursuit. In December 2022, the South Korean military failed to track five North Korean drones that invaded South Korean airspace, sending them all back to North Korea \cite{north}. In modern warfare, drone attacks while operating loitering munitions have become an inevitable strategy to gain the upper hand in wars. As one example of the use of drones in wars, military-purpose drones are also used in the current war between Ukraine and Russia. The SwitchBlade 300 loitering munition, made by AeroVironment of the United States, is being used successfully to attack the Russian garrison and tanks~\cite{ukraine}. According to the Russian defense ministry, the drone attack on Russian bases fatally wounded three technical Russian staff~\cite{ukraine3}. Furthermore, dogfights between reconnaissance drones have also occurred \cite{ukraine2}.
There have also been some cases of massive damage caused by drone attacks. Saudi Arabia's state-owned oil refinery enterprise, Aramco, was hit by more than 17 loitering munition trials at its two refineries in 2019~\cite{aramco}. More than 20 drones each carried more than three kilograms of explosives and suffered massive damage, disrupting supplies of 5.7 million barrels of petroleum. It was a situation caused by the lack of defense measures and an opportunity to remind people of the need to build defense systems.




\subsection{Algorithm Design Rationale}
According to the modern trends in the usage scenarios of drone networks, we can observe that drones can be effective and efficient in battlefield autonomous unmanned defense system design based on autonomous multi-drone UAV mobility control cooperation and coordination. For the purpose, deep reinforcement learning (DRL) algorithms can be the best solution due to the nature of discrete-time stochastic control and sequential decision-making processes. Among various DRL algorithms, our proposed algorithm is fundamentally designed based on deep deterministic policy gradient (DDPG) for utilizing continuous action space in order to realize nonlinear drone mobility control~\cite{iotj20kwon, 9685905,globecom19kwon}. Furthermore, for dealing with unexpected blockage and real-time environment changes dynamically in the physical worlds, situation-aware mechanisms are additionally considered on top of DDPG-based DRL design. In previous research results, drones fly over slight-line trajectories, thus it is not possible to consider nonlinear drone mobility control for avoiding blockages~\cite{9685905,ref1,infocom21lee,9337915,8869876}.

\subsection{Contributions} 
The major contributions of this research in situation-aware DDPG-based nonlinear drone mobility control can be summarized as follows.

\begin{itemize}
    \item First of all, our proposed DRL-based algorithm fundamentally aims at autonomous nonlinear drone mobility control based on near-field situation sensing for taking care of obstacles. In the literature, many approaches are using pure DRL algorithms without situation-awareness, which can be harmful in an urban battlefield environment where the locations are with many obstacles (blockages, buildings, and structures). This situation-aware nonlinear mobility control is definitely beneficial in our considering applications, i.e., autonomous drone mobility control for loitering munition. 
    \item Moreover, the proposed situation-aware DRL-based autonomous nonlinear drone mobility control algorithm is also implemented with Unity 3D Environment for loitering munition mobility control. This implementation with Unity 3D Environment is the essential part for the realization of CPS for interacting and matting the actions in the virtual environment and the ones in the real world for better intuitive and efficient understanding to the algorithm designers and system users (military officers and soldiers). 
    \item Lastly, compared to our previous work \cite{APWCS22lee}, this paper mathematically specified the algorithm and the performances are evaluated in various ways. 
\end{itemize}

\subsection{Organization}
The rest of this paper is organized as follows. Sec.~\ref{sec:related} presents the related work of drone technologies and reinforcement learning. Sec.~\ref{sec:algorithm} explains our proposed situation-aware autonomous nonlinear drone mobility control algorithm with mathematical analysis, and Sec.~\ref{sec:performance} includes simulation-based performance evaluation results with corresponding discussions. Finally, Sec.~\ref{sec:conclusion} concludes this paper and presents future research directions.

\section{Related Work}\label{sec:related}

There are a lot of research results in terms of drone trajectory optimization and mobility control. In order to design the drone mobility control algorithms, each research contribution is with its own objective for the mobility control. 
In~\cite{icoin21jung}, DRL-based multi-drone mobility control algorithms are designed and implemented under the considerations of weights and batteries which can restrict the drone's flight time and mobility. 
This type of research results is to optimize the trajectories of drones to move efficiently within a limited time~\cite{twc20lee}. 
In addition, free-space optical communication (FSOC) is employed to deal with the trajectory optimization of a fixed-wing UAV over various atmospheric conditions, as well-presented in~\cite{twc20lee}.

The autonomous drone mobility control algorithms can be also designed under the consideration of the drone's specific applications and objectives. In~\cite{tii22yun}, the drone mobility control algorithms are designed using multi-agent DRL for CCTV-based surveillance systems in smart city applications. In addition, autonomous drone mobility control algorithms can be designed for cooperative and coordinated mobile access points and base-stations positioning~\cite{9845684,9824676,8660516,9525079}. 
Furthermore, the proposed DRL-based algorithm in~\cite{icte21yun} aims at optimal passenger delivery in urban air mobility (UAM) applications (i.e., drone-taxi) while avoiding accidents. The algorithm in~\cite{icte21yun} is distributed DRL-based mobility control to electric vertical takeoff and landing (eVTOL) drone platforms for UAM passenger delivery applications. The eVTOL vehicles search for the optimal passenger transport routes under the consideration of passengers' behaviors, collision potentials, and battery status levels through QMIX-based multi-agent DRL.




According to the fact that it is essential to develop power-charging algorithms in power-hungry drones, the proposed multi-agent DRL algorithm in~\cite{tvt201905shin} studied an outdoor environment ad-hoc battery charging method for multiple drone environment. For the sake of overcoming the lack of charging platform and the pressing of charging time, an auction-based resource allocation using deep learning framework is employed to perform the charging schedule. 
The main reason why auction-based algorithms should be used in this problem is that it is fully distributed (i.e., also working with only neighbor information under uncertainty) and truthful~\cite{tvt201905shin}.
The proposed Myerson auction with a deep learning framework is to ensure dominant strategy incentive compatibility (DSIC) and individual rationality (IR) for truthful operations. According to the data-intensive performance evaluation for the proposed algorithm, it ensures optimal revenue under the considerations of DSIC and IR while achieving revenue-optimality.
Lastly, the proposed multi-agent DRL-based drone mobility control algorithm in~\cite{tvt202106jung} is for joint cooperative power-charging control and drone charging scheduling under the support of built-in infrastructure, i.e., charging towers. The proposed algorithm in the paper determines scheduling between drones and charging towers at first. During this phase, it has been confirmed that the proposed scheduling optimization framework is non-convex; thus, it is converted to convex with some mathematical theories. After that, the amounts of power-charging in the scheduled drone-tower pairs are determined based on multi-agent DRL. }

\section{Situation-Aware and DRL-based Autonomous Nonlinear Drone Mobility Control for Loitering Munition}\label{sec:algorithm}

\begin{figure}[t!]
    \centering
     \includegraphics[width =0.99\linewidth]{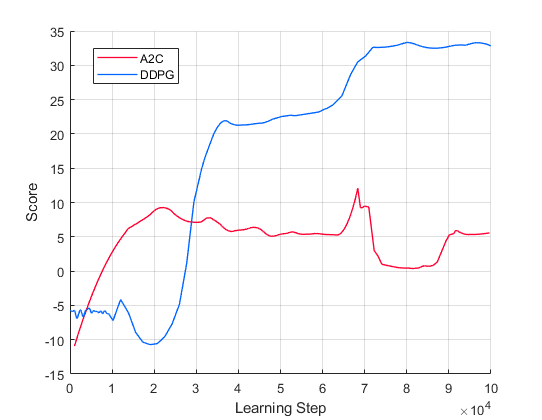}
    \caption{RL score plotting (i.e., reward value dynamics) for each step.}
    	\label{fig:RL_score}
\end{figure}

{
Our proposed autonomous nonlinear drone mobility control algorithm is fundamentally designed on top of the deep deterministic policy gradient (DDPG) algorithm, which is one of the well-known policy-gradient (PG) DRL frameworks. By utilizing the PG-based DRL algorithms, the shortcomings of the existing value-based deep Q-network (DQN) algorithms can be compensated~\cite{iclr16lillicrap}. Unlike the value-based algorithms, which can only be applied to the environments with discrete actions, DDPG-based DRL algorithms can handle the continuous movements of drones~\cite{iotj20kwon}. 
To conduct action inference computation in A2C, an $\arg\max$ function should be used after passing through the actor network, which is infeasible for continuous action spaces. 
To take care of this issue, our proposed autonomous nonlinear drone mobility control algorithm is designed based on DDPG because it aims at continuous action control. There are also differences in how the network is updated for training via gradient descent. A2C updates its own actor network by evaluating actions, using the estimation of its own Q-function. On the other hand, our considering DDPG-based algorithm updates the critic network by temporal difference (TD) target to improve the stability.

When the action control and experiments are conducted for drone aerial movements using conventional PG-based DRL algorithms such as actor-critic (A2C)-based algorithms, the corresponding modeling can increase (+) and decrease (-) the movements in 3D $x$, $y$, and $z$ directions, e.g., $(1, 0, 0)$, $(-1, 0, 0)$, $(0, 1, 0)$, $(0,-1, 0)$, $(0, 0, 1)$, and $(0, 0, -1)$, respectively. However, according to the simulation-based verification as shown in Fig.~\ref{fig:RL_score}, we can observe that the performance of A2C-based algorithms is significantly lower than that of DDPG-based algorithms because DDPG-based algorithms can control the mobility via continuous action control which is easily tractable for nonlinear continuous behavior modeling and computation. 

Note that the parameters of the actor neural network and the critic neural network as denoted as $\theta^\mu$ and $\theta^Q$ in our desired DDPG-based DRL algorithm computation. In addition, the objectives of actor $\mu(\cdot|\theta^\mu)$ and critic network $Q_{\theta^Q}$ are for our actor network to maximize action-value derived from critic network, as well as for our critic network to follow Bellman's equation which consists of the target critic network $Q_{\theta^{\hat{Q}}}$ and reward function, respectively. For this purpose, our proposed DDPG-based DRL algorithm adopts a greedy policy that the actor network approximates the action which maximizes action-value function (i.e., critic network) for given state $\mathbf{x}_t$, which can be expressed as,
\begin{equation}\label{eq:ddpg1}
    \mu(\mathbf{x}_t|\theta^\mu) \approx \arg\max_{\mathbf{u}_t} Q_{\theta^Q} (\mathbf{x}_t, \mathbf{u}_t).
\end{equation}

The parameter of the actor neural network should meet the following optimization criteria,
\begin{equation}\label{eq:ddpg2}
    {\theta^\mu}^* =  \arg\max_{\theta^\mu} Q_{\theta^Q} (\mathbf{x}_t, \mu(\mathbf{x}_t|\theta^\mu))
\end{equation}
and the parameter $\theta^\mu$ can be calculated by maximizing $Q_{\theta^Q}$ based on following stochastic gradient ascent computation,
\begin{equation}\label{eq:ddpg3}
    \theta^\mu \leftarrow \theta^\mu + \alpha \nabla_{\theta^\mu} Q_{\theta^Q} (\mathbf{x}_t, \mu(\mathbf{x}_t|\theta^\mu)).
\end{equation}

Now, the gradient for $\theta^\mu$ in $Q_{\theta^Q}$ can be expressed as follows,
\begin{align}
  \nabla_{\theta^\mu} Q_{\theta^Q} &= \frac{d Q_{\theta^Q}}{d\theta^\mu} \\
  & = \frac{d\mathbf{u}_t}{d\theta^\mu} \frac{d Q_{\theta^Q}}{d\mathbf{u}_t} \\ 
  &= \nabla_{\theta^\mu} \mu(\mathbf{x}_t|{\theta^\mu})\nabla_{\mathbf{u}_{t}} Q_{\theta^Q}  (\mathbf{x}_t, \mathbf{u}_t) \label{eq:ddpg4}
\end{align}
according to a chain rule.

We can now calculate TD error $y_i$ using the target critic neural network as follows,
\begin{align}\label{eq:ddpg5}
  y_i &= r(\mathbf{x}_i, \mathbf{u}_i) + \gamma \max_{\mathbf{u}_{i}'} Q_{\theta^{\hat{Q}}}(\mathbf{x}_{i}',\mathbf{u}_{i}') \\
  &= r(\mathbf{x}_i, \mathbf{u}_i) + \gamma Q_{\theta^{\hat{Q}}}(\mathbf{x}_{i}', \arg\max_{u_{i+1}} Q_{\theta^{\hat{Q}}}(\mathbf{x}_{i}', \mathbf{u}_{i}')) \\
  &\approx r(\mathbf{x}_i, \mathbf{u}_i) + \gamma Q_{\theta^{\hat{Q}}}  (\mathbf{x}_{i}', \mu(\mathbf{x}_{i}' | {\theta^{\hat{\mu}}}))
\end{align}
where $\mathbf{x}_{i}'$ and $\mathbf{u}_{i}'$ stand for the next state and the action given state-action $(\mathbf{x}_i, \mathbf{a}_i)$, respectively.
The loss function of actor neural network is defined as follows, 
\begin{equation}\label{eq:ddpg6}
    L({\theta^\mu}) = - \sum_{\forall i} Q_{\theta^Q} (\mathbf{x}_i, \mu(\mathbf{x}_i | {\theta^\mu}))
\end{equation}
where 
\begin{equation}
\nabla_{\theta^\mu} Q_{\theta^{Q}} (\mathbf{x}_i , \mu(\mathbf{x}_i | {\theta^\mu}))
\end{equation}
is the gradient obtained by stochastic gradient descent methods, and thus, \eqref{eq:ddpg6} indicates that the neural network updates its parameters for the direction of loss function minimization. 

In addition, the loss function of the critic neural network is calculated with a TD target as follows,
\begin{equation}
\label{eq:ddpg7}
    L(\theta^\mu) = \frac{1}{2} \sum_{\forall i}  ||y_i - Q_{\theta^Q}(\mathbf{x}_i,  \mu(\mathbf{x}_i|{\theta^\mu}))||^2,
\end{equation}
where 
\begin{equation}
y_i = r(\mathbf{x}_i, \mathbf{u}_i) + \gamma Q_{\theta^{\hat{Q}}}  (\mathbf{x}_{i}', \mu(\mathbf{x}_{i}' | {\theta^{\hat{\mu}}})), 
\end{equation}
as described in \eqref{eq:ddpg5}. 
Note that the action $\mu(\mathbf{x}_{i}' | {\theta^{\hat{\mu}}})$ is calculated by applying target actor neural network parameterized to $\theta^{\hat{\mu}}$. Our proposed DDPG-based DRL algorithm applies soft target update method, which means the parameter of the target neural network follows the actor neural network slowly, setting $\gamma$ with a very small value, i.e.,
\begin{eqnarray}\label{eq:ddpg8}
  \theta^{\hat{Q}} &\leftarrow& \gamma \theta^Q + (1-\gamma)\theta^{\hat{Q}} \nonumber \\
  \theta^{\hat{\mu}} &\leftarrow& \gamma \theta^{\mu} + (1-\gamma)\theta^{\hat{\mu}}.
\end{eqnarray}

According to the fact that our proposed DDPG-based DRL algorithm is a deterministic policy based algorithm, it is required to provide randomness to the action. Therefore, a noise factor $\epsilon_t$ is added to the action as shown in \eqref{eq:ddpg9},
\begin{equation}\label{eq:ddpg9}
    \mu_{\textit{noisy}} (\mathbf{x}_t) = \mu(\mathbf{x}_t | \theta^\mu) + \epsilon_t.
\end{equation}

Our proposed algorithm's state space contains information about the agent and the environment. The difference between the coordinate vector of the drone and the target, and the vector represents the drone's flight speed and angular speeds are included in the state space. For realizing the obstacle factors in drone mobility control, a Raycast function is provided by Unity in the state space. Note that the Raycast is a function that shoots a virtual laser, and it can detect the direction and distance between the agent and the obstacle by shooting lasers in vertical or horizontal directions. In addition, the actions are defined as the moves over $x$, $y$, and $z$ directions, respectively. Finally, we can confirm that situation-awareness can be realized by the Raycast function for utilizing virtual laser functionalities.
\begin{figure}[t!]
    \centering
    \includegraphics[width =0.99\linewidth]{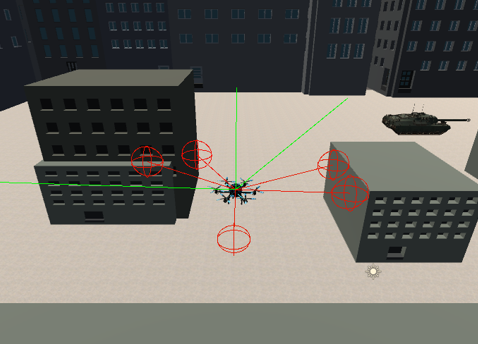}
    \caption{Raycast information for situation-awareness which is essentially required for autonomous nonlinear drone mobility control.}
    	\label{fig:Raycast}
\end{figure}
The proposed algorithm is different from the existing conventional DDPG because it can recognize surrounding objects, i.e., situation-awareness. As explained, we added the Raycast function provided by Unity to identify nearby obstacles, where the Raycast is a function that casts virtual rays from the moving object to measure the distance to another object or determine whether the laser reaches an object. Fig.~\ref{fig:Raycast} represents the Raycast shot from the agent. In Fig.~\ref{fig:Raycast}, the drone/agent can know that there are no obstacles at the 12 o'clock, 1 o'clock, and 9 o'clock directions through the Raycast. In addition, considering the previously acquired location information of the target, the drone/agent will act to fly in the direction of 1 o'clock. A ray in the vertical direction measures the distance to the ground and is used to avoid colliding with the ground. In order to induce the drone/agent to avoid surrounding obstacles, when the launched ray collides with an obstacle, the drone/agent gets a negative reward that is inversely proportional to the distance to the obstacle. By recognizing obstacles through Raycast, our proposed situation-aware autonomous nonlinear drone mobility control algorithm is designed. 


\begin{algorithm}[t]
	Initialize the \textit{critic} and \textit{actor} networks of the drone agent with weights $\theta^{Q}$ and $\theta^{\mu}$~\\
	Initialize the target networks as: $\theta^{\hat{Q}} \leftarrow \theta^{Q}, \theta^{\hat{\mu}} \leftarrow \theta^{\mu}$ ~\\
	\For{episode = 1, $\mathcal{E}$}{
		\textbf{Phase 1. Initialize the replay buffer $\mathcal{D}$}~\\ 
		\For{mini batch = 1 to c}{
			$\triangleright$ Randomly generate $\varphi$ states $\textbf{x} \in \mathbb{X}$ ~\\
			$\triangleright$ \textbf{Observe surrounding obstacles through Raycast} and store to the states $\textbf{x}$ to aware situation\\   
			$\triangleright$ Get corresponding a set of actions $\textbf{u} = \mu(\textbf{x}| \theta^{\hat{\mu}}) \in \mathbb{U}$ for each $\textbf{x}$~\\
			$\triangleright$ Input the state-action pairs to predefined \textbf{drone environments} and get a set of reward $r$ for each pair, and observe the next set of states $\textbf{x}' \in \mathbb{X}'$}
		$\triangleright$ Store the transition pairs $\xi = (\textbf{x}, \textbf{u}, r, \textbf{x}')$ as a minibatch, which composes $\mathcal{D}$.
	} 
	\textbf{Phase 2. Update neural networks periodically~\\}
	\For{time step = 1, $\mathcal{T}$}{
		\textbf{If} \textit{time step} \textbf{is update period, do followings:}~\\
		$\triangleright$ Sample a random minibatch $\mathcal{V}$ without replacement from $\mathcal{D}$~\\
		$\triangleright$ Set $y_{i} = r_{i} + \gamma \hat{Q}(\textbf{x}_{i}', \mu(\textbf{x}_{i}'|\theta^{\hat{\mu}})|\theta^{\hat{Q}})$ ~\\
		$\triangleright$ Update the $\theta^{Q}$ by applying stochastic gradient descent to the loss function of \textit{critic} network, which can be obtained as $\frac{1}{\mathcal{V}}\sum_{i}{(y_{i} - Q(\textbf{x}_{i}, \textbf{u}_{i}|\theta^{Q}))}^{2}$ ~\\
		$\triangleright$ Update the $\theta^{\mu}$ by applying stochastic gradient ascent concerning the gradient of \textit{actor} network: ~\\
		$\nabla_{\theta^{\mathcal{U}}}J(\theta^{\mu}) \approx \frac{1}{\varphi}\sum_{i}{\nabla_{\textbf{u}}Q(\textbf{x}, \textbf{u}|\theta^{Q})\nabla_{\theta^{\mu}}\mu(\textbf{x}|\theta^{\mu})|_{\textbf{x}=\textbf{x}_{i},\textbf{u}=\mu(\textbf{x}_{i}|\theta^{\mu})}}$~\\
		$\triangleright$ \textit{Soft} update $\theta^{\hat{Q}}$ and $\theta^{\hat{\mu}}$ as follows:
		$\theta^{\hat{Q}} \leftarrow \gamma \theta^Q + (1-\gamma)\theta^ {\hat{Q}}$, 
		$\theta^{\hat{\mu}} \leftarrow \gamma \theta^{\mu} + (1-\gamma)\theta^ {\hat{\mu}}$
		}
	\caption{Proposed situation-aware autonomous nonlinear drone mobility control algorithm}
	\label{algorithm1}
\end{algorithm}

The pseudo-code of our proposed situation-aware DDPG-based algorithm is described in Algorithm~\ref{algorithm1}. 
First of all, the weights $\theta^Q$ and $\theta^\mu$ of actor/critic networks and the weights $\theta^{\hat{Q}}$ and $\theta^ {\hat{\mu}}$ of target network are initialized. 
After that, a replay buffer is also initialized in the last part of the episode. For each minibatch, $\varphi$ states are randomly generated and get the appropriate set of actions $\textbf{u} \in \mathbb{U}$ for each state $\mathbf{x}$. After that, the drone/agent recognizes obstacles around itself through Raycast observation for situation-aware DRL computation, and stores the information in state space $\mathbf{x}$, in (line 7). Then we input the $\mathbf{x}$ and $\mathbf{u}$ pairs to predesigned drone environments and obtain the set of reward $r$ for each pair, and observe the next state $\mathbf{x}'$. The actor network uses the state obtained from the drone as an input, and calculates an appropriate action as an output through two fully connected layers. The drone agent that receives the action $\mathbf{u}$ performs the corresponding action in the environment and gets the next state $\mathbf{x}'$. The observed transition pairs $\xi$ from the drone/agent are saved as a minibatch $\mathcal{V}$, as shown in Fig.~\ref{fig:envtobuffer}. 

\begin{figure}[t!]
    \centering
    \includegraphics[width =0.98\linewidth]{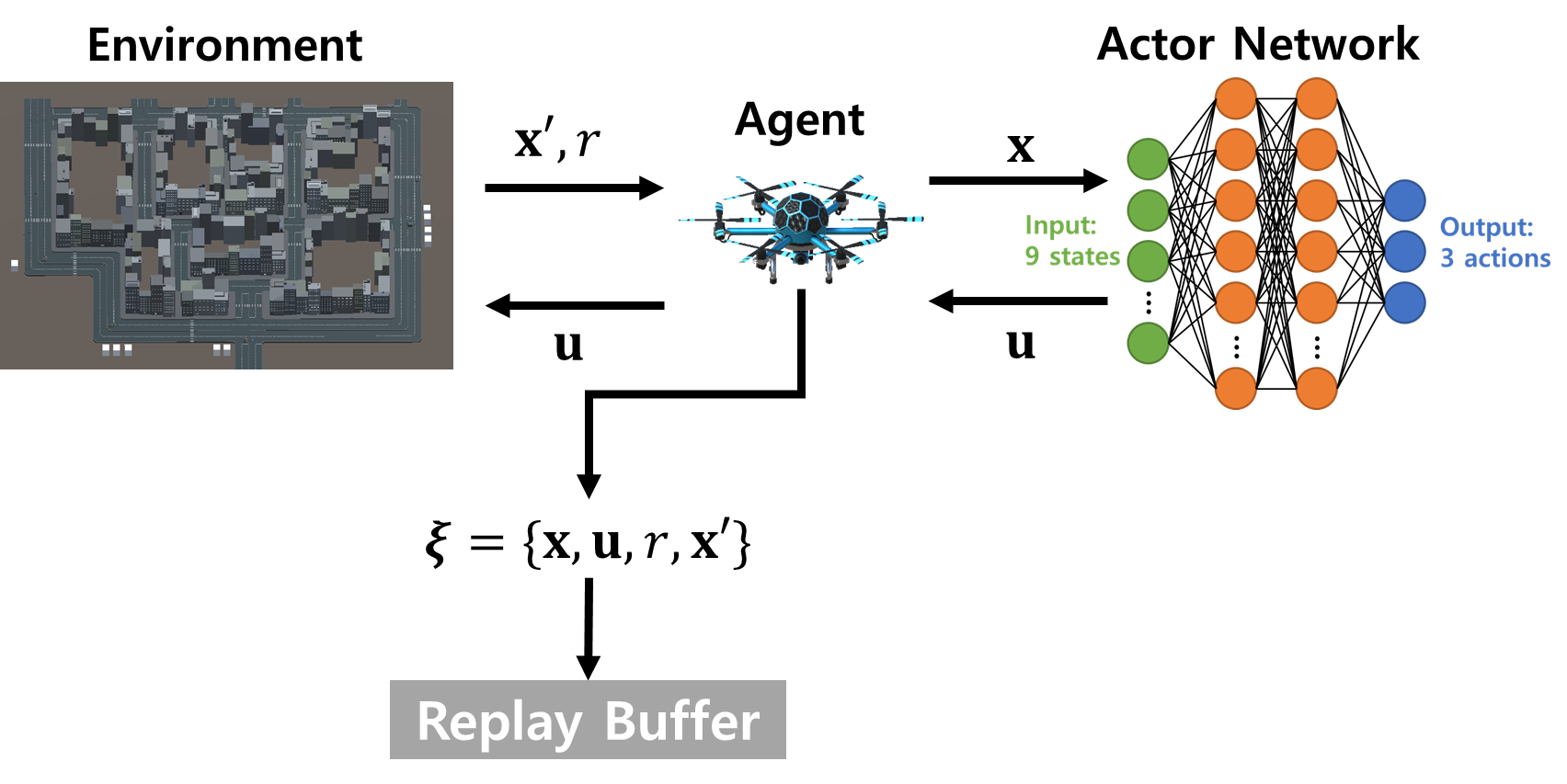}
    \caption{The information from the drone/agent's observation is stored in a replay buffer.}
    	\label{fig:envtobuffer}
\end{figure}

\begin{figure*}[t!]
    \centering
    \includegraphics[width =0.68\linewidth]{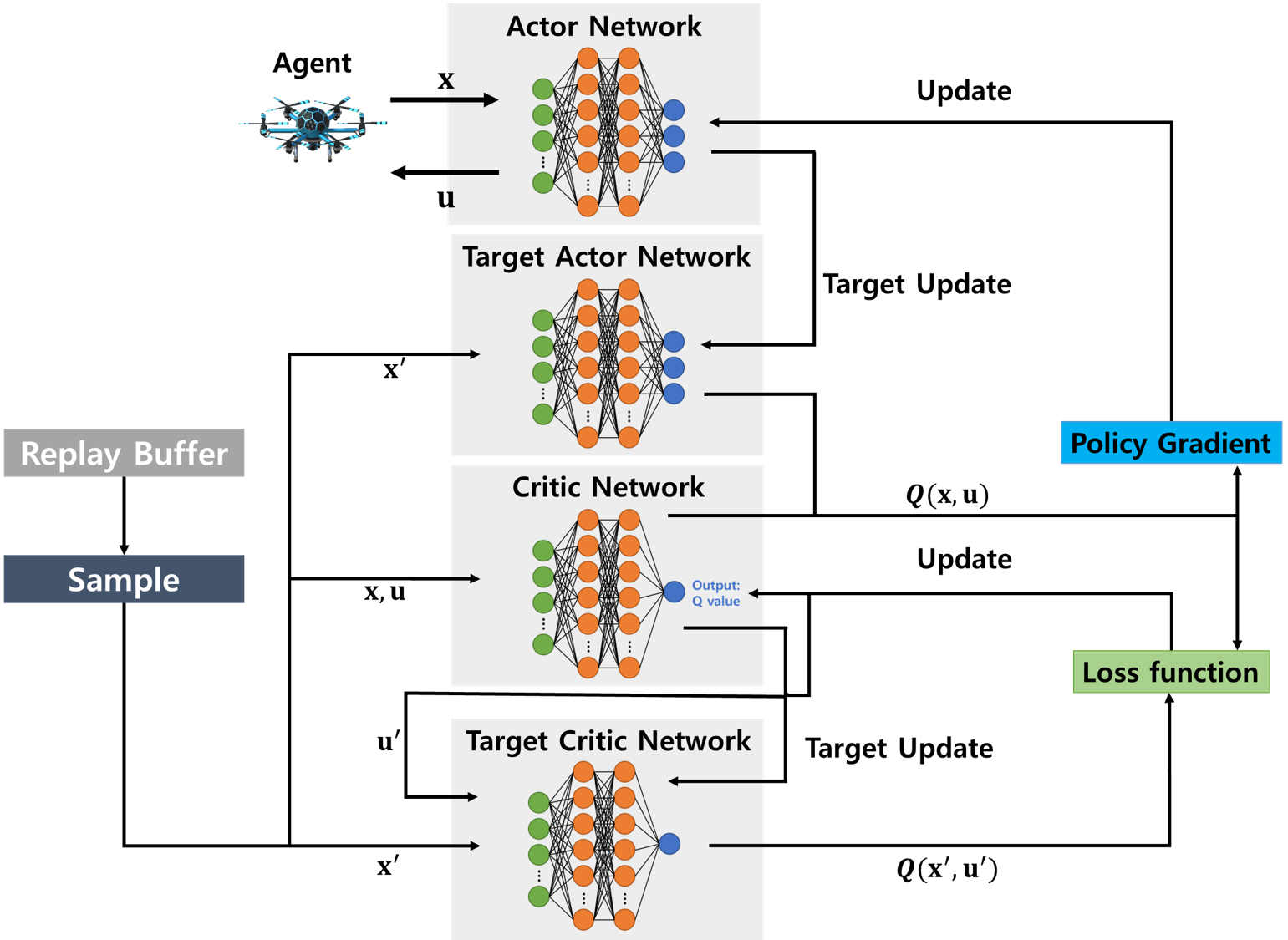}
    \caption{Overall computation process of neural network parameter updates in DDPG-based DRL algorithms.}
    	\label{fig:DDPG_process}
\end{figure*}

During the next phase, we should update the networks regularly. For each time step, if the time step is in the update period, we sample a random minibatch $\mathcal{V}$ without replacement from the replay buffer $\mathcal{D}$. In the successive policy decision process, in order to solve the correlation problem between samples, the learning outcome can be increased by performing learning in the replay buffer in the unit of samples. To obtain the network's TD target, this algorithm calculates the equation in (line 17) of Algorithm~\ref{algorithm1}. The random sample extracted from the tuple delivered to the replay buffer by the drone/agent in Fig.~\ref{fig:envtobuffer} becomes the input of the target actor network, critic network, and target critic network, as shown in Fig.~\ref{fig:DDPG_process}. Finally, this algorithm updates the target critic and the target actor network for $\mathcal{T}$ time steps and ends the algorithm computation procedure. The stability of learning can be increased through the usage of target networks. The update of the critic network and the two target networks using the loss function and the policy gradient method is a series of processes for maximizing the expected reward by eventually deriving more appropriate actions for the actor network.

\section{Performance Evaluation}\label{sec:performance}
\begin{figure}[t!]
    \centering
    \includegraphics[width =0.98\linewidth]{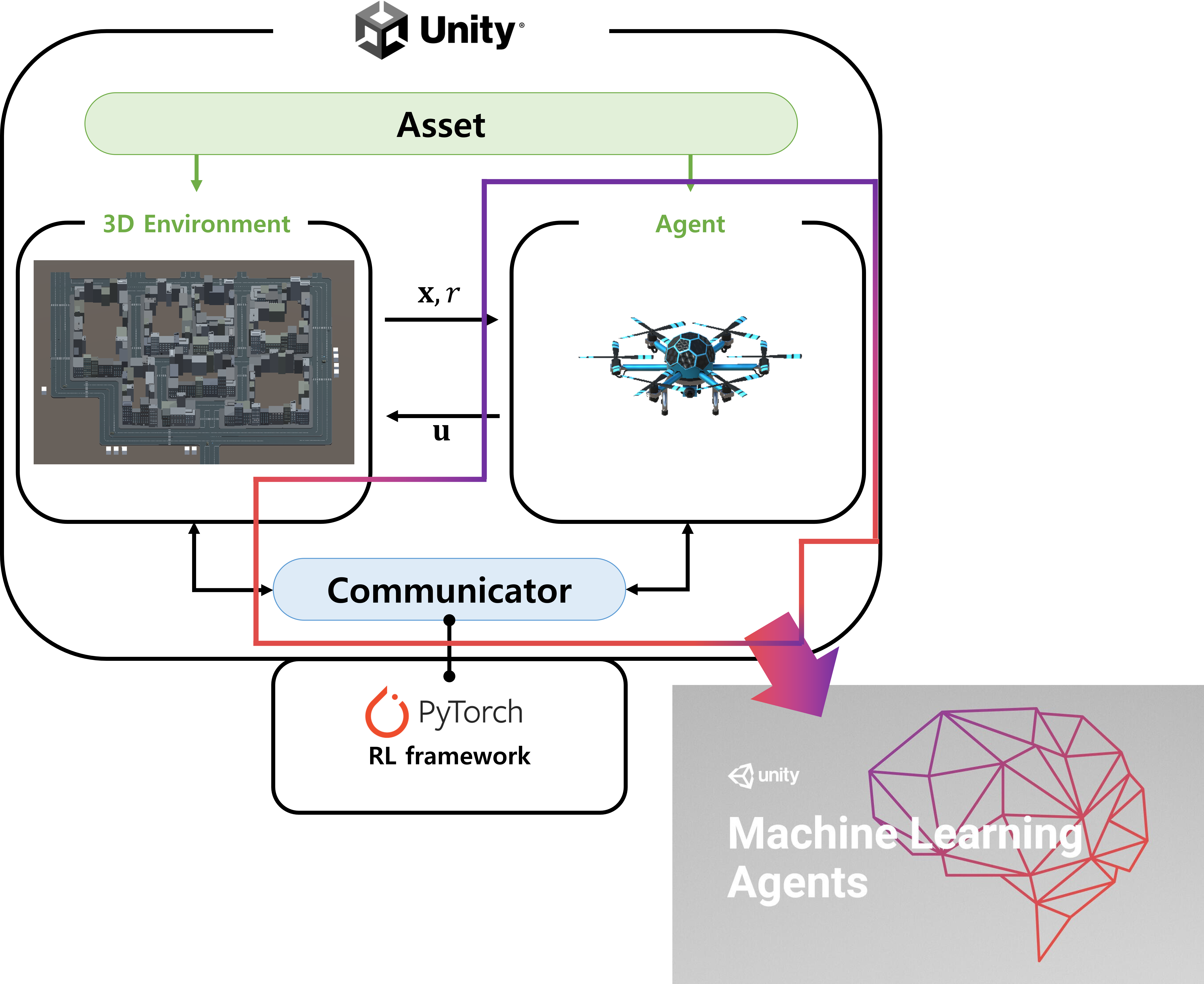}
    \caption{Software architecture in Unity environment.}
    	\label{fig:software}
\end{figure}

\subsection{Unity Environment}\label{sec:system}
Unity is a 3D tool-creating platform that utilizes Unity Machine Learning Agents (ML-Agents), an API to support artificial intelligence for training agents through Unity. 
The ML-Agents basically support various functions for DRL learning computation.
In real world scenarios, the use of DRL algorithms to multi-drone mobility control takes much longer time and can lead to costs, tens of times or more. Therefore, the use of DRL algorithms in a simulation environment is suitable, whereas it is not suitable in the physical/real world. The software architecture for RL/DRL trainers and Unity environment is illustrated in Fig.~\ref{fig:software}. We build the environment using the assets furnished by Unity 3D and set it by applying the ML-Agent to the environment \cite{unitymlagents}. Then the drone/agent is trained by employing a PyTorch-based reinforcement learning trainer. When the learning is completed, it is embedded into the Unity environment through the communicator. Finally, the Unity environment with the trained agent is created. 

\subsection{Evaluation Settings}
\begin{figure*}[h!]
    \centering
    \includegraphics[width =0.7\linewidth]{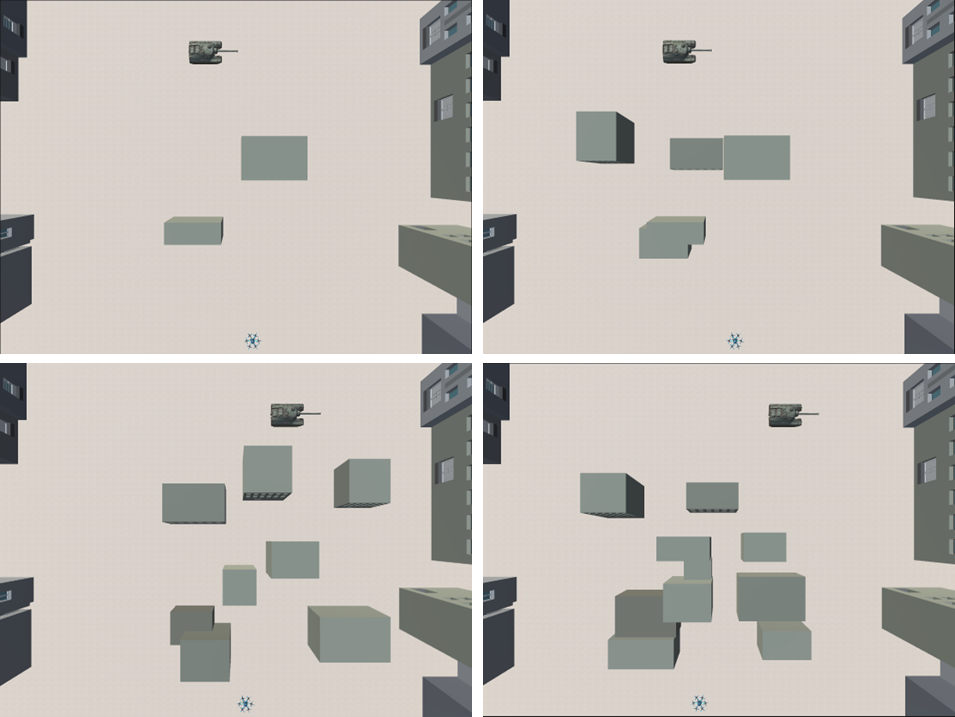}
    \caption{Top-down view of the environment where the obstacle density is set to 20\%, 50\%, 80\%, and 100\%, respectively.}
    	\label{fig:Density}
     \vspace{3mm}
\end{figure*}
We build the city environment (for emulating urban military battlefield scenarios) through the Unity Asset Store. In addition, we add drone and tank assets to the environment scene. The drone is set as an agent, and the target with the tank shape is set as a game object. The batch size of each neural network is set to 128, and the discount factor from the previous step is set to 0.9. The learning computation is carried out for a total of 100,000 steps, and the test step is performed 10,000 times. The critic network updates the Q-function in the direction of reducing the difference between the predicted value and target value. Moreover, the policy of the actor network is updated to maximize the objective function through two fully connected layers, which have the size of 128. The agents and targets are generated randomly within a designated area, respectively, and the target moves forward at a constant speed. To ensure that the agent does not consider staying in place as optimal, we start with the addition of the reward of -0.01 by default. In order to induce the drone to approach the target gradually, a value obtained by subtracting the current distance from the previous distance is reflected in the compensation function. When the drone reaches the target, it receives a reward of +1; if it is too far from the target or collides with the obstacle, it receives a reward of -1 and ends the episode. If an obstacle is recognized nearby through the Raycast (i.e., situation-awareness), a negative reward is given so that the agent learns to avoid the obstacle. By acquiring information about surrounding obstacles, the proposed situation-aware autonomous nonlinear drone mobility control algorithm can obtain better results compared to the conventional DDPG-based DRL algorithm.

In order to evaluate the algorithm in various ways and more difficult environments, the cyber environment that can be harsher (i.e., more obstacles) than the physical situation is constructed. By setting different densities of obstacles in the cyber environment with no obstacles at all to environments where it is harsh for the agent to pass through, the robustness of the algorithm can be evaluated, especially in environments with densely distributed obstacles. Fig.~\ref{fig:Density} depicts the top-down view of the environment where the obstacle density is set to 20\%, 50\%, 80\%, and 100\%, respectively.
Note that the obstacles are created in random sizes in random locations between the agent and the target in each episode. The agent and the target are also produced at random positions within a designated space.


\begin{figure}[h!]
    \centering
    \includegraphics[width =0.99\linewidth]{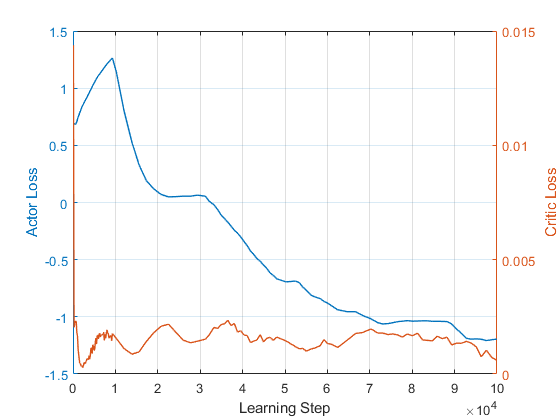}
    \caption{The actor and critic loss value dynamics of our proposed DDPG-based nonlinear control algorithm in each episode.}
    	\label{fig:Losses}
\end{figure}


\subsection{Evaluation Results}

\begin{figure}[t!]
    \centering
    \includegraphics[width =0.9965\linewidth]{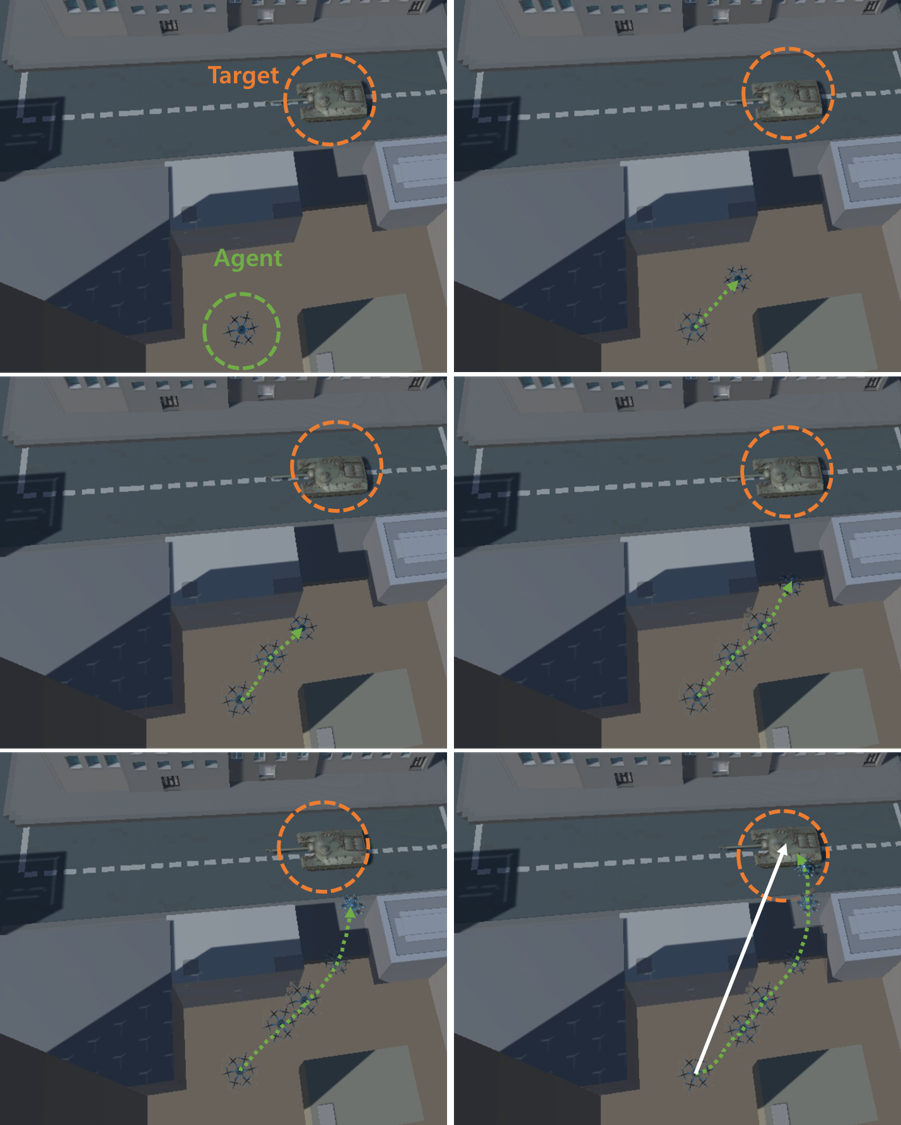}
    \caption{Top-down view for the drone/agent approaching the target while avoiding obstacles.}
    	\label{fig:6p_TD_3x2}
\end{figure}

\begin{figure}[t!]
    \centering
    \includegraphics[width =0.9965\linewidth]{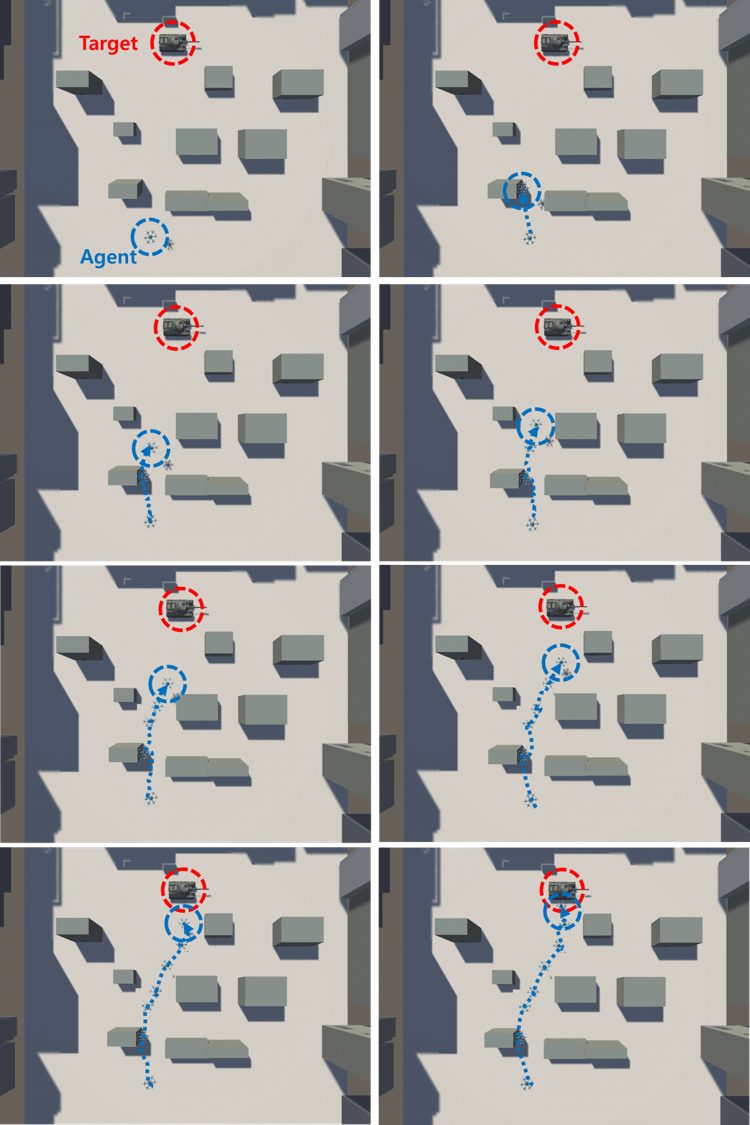}
    \caption{Top-down view for the drone/agent approaching the target while avoiding obstacles (obstacles density: 90\%).}
    	\label{fig:8p_TD_4x2}
\end{figure}

\begin{figure}[h!]
    \centering
    \includegraphics[width =0.9965\linewidth]{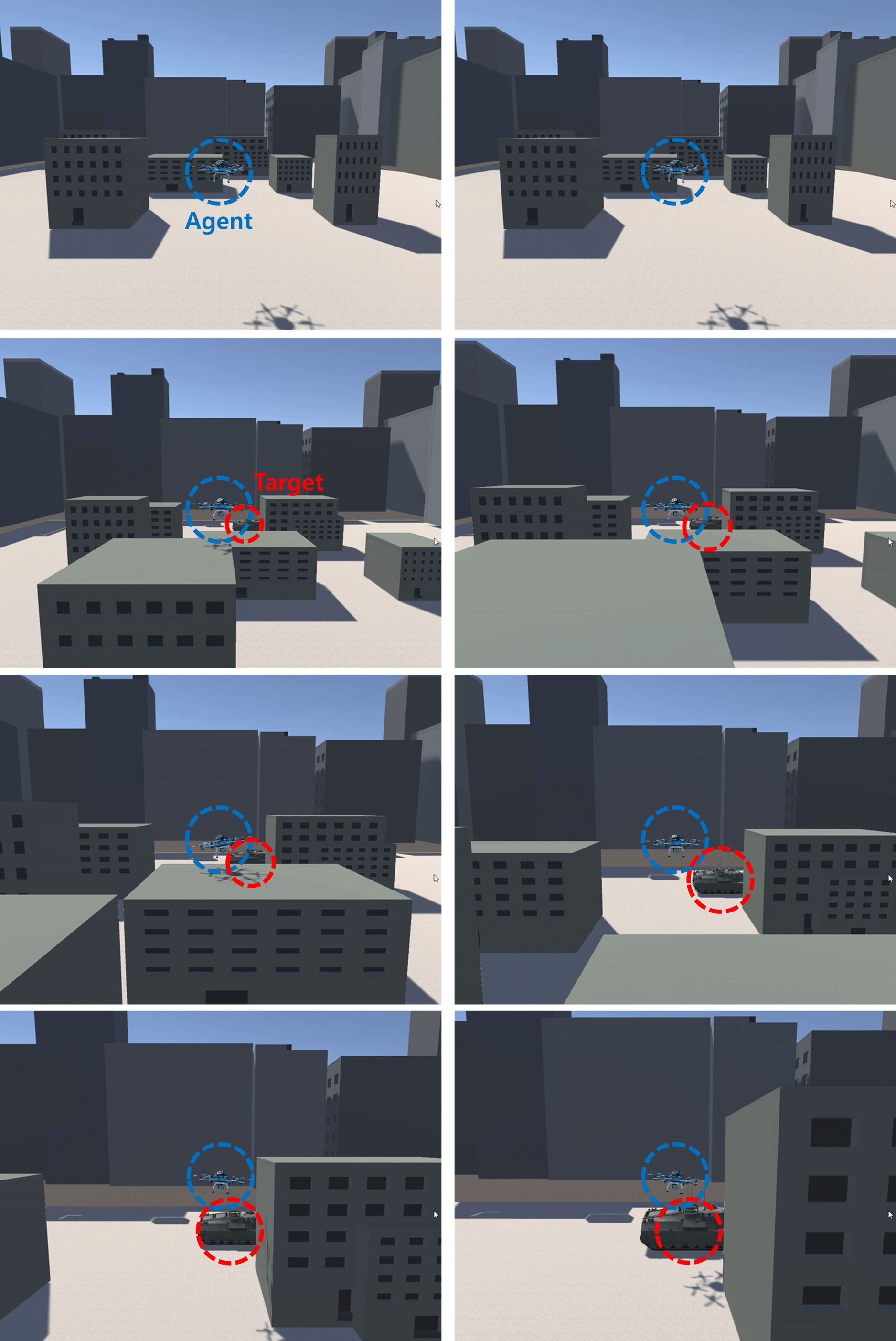}
    \caption{Third-person view for the drone/agent approaching the target while avoiding obstacles (obstacles density: 90\%).}
    	\label{fig:8p_TP_4x2}
\end{figure}

Fig.~\ref{fig:Losses} shows the actor loss and critic loss, which indicates the convergence of the proposed actor/critic-enabled DDPG-based DRL algorithm. The actor loss and critic loss values also decreased as learning data were accumulated, as described in Fig.~\ref{fig:Losses}. A decrease in actor loss means that the action value is maximized, and thus the cumulative reward is maximized. In addition, the reduction in the critical loss indicates that the actor network approaches the optimal Q-network according to Bellman's expectation equation, as described in~\eqref{eq:ddpg5}.
Fig.~\ref{fig:6p_TD_3x2} depicts the case where a drone reaches a target (tank) while avoiding obstacles in six frames, in a physical urban environment. As shown in Fig.~\ref{fig:6p_TD_3x2}, a trained drone/agent can accurately reach a target (tank). The white arrow is a predicted trajectory of the conventional linear mobility control (LMC), which may not be able to reach the target. It can be seen that the learning was performed smoothly in both the visualized simulations and the corresponding numerical values.

In order to show the robustness of the algorithm, the experimental results in the cyber-physical space with many obstacles are shown in Fig.~\ref{fig:8p_TD_4x2} and Fig.~\ref{fig:8p_TP_4x2}. According to the benefits of the nature of autonomous nonlinear drone mobility control, the drone/agent found its destination well, even in the cyber CPS systems with dense obstacles. In Fig.~\ref{fig:8p_TD_4x2}, we can find out that the drone/agent controls its altitude autonomously, as displayed through the drone/agent's shadow. We can check the trajectory drawn by the drone/agent through the top-down view, but we cannot confirm the height difference of the buildings. For more details, in Fig.~\ref{fig:8p_TP_4x2}, we can see that the drone/agent is controlling its altitude autonomously, as depicted in a third-person view avoiding obstacles and reaching the target/tank. In the top-down view, it may seem to fly very close to the building, but in the third-person view expressed in Fig~\ref{fig:8p_TP_4x2}, you can confirm that the drone is flying while adjusting the altitude and distance from the building.

\begin{figure*}[t]\centering
    \begin{multicols}{2}
    \includegraphics[width=0.99\linewidth]{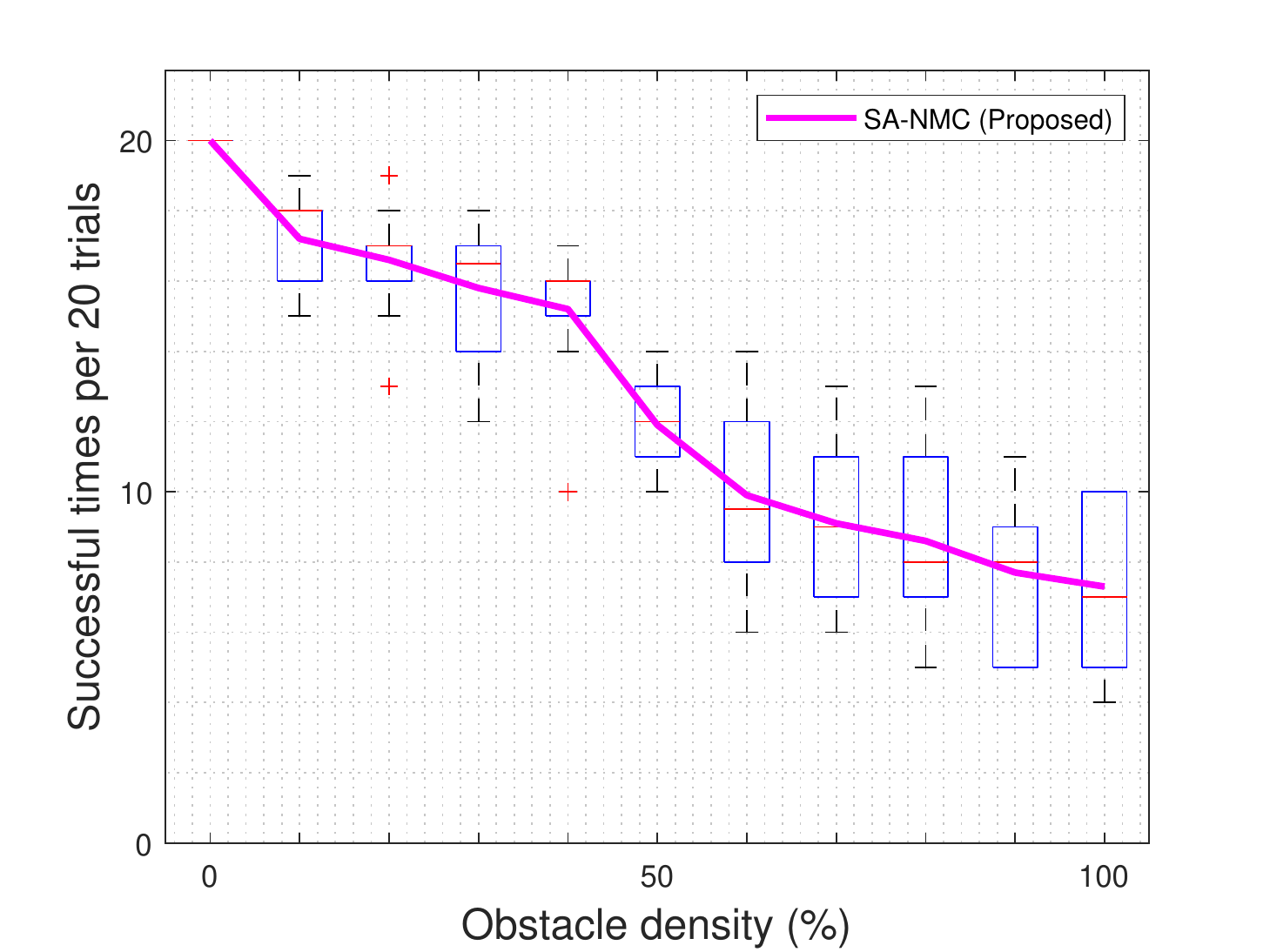}\captionsetup{justification=centering}\\
    \subcaption{\textbf{Proposed}}
    
     \includegraphics[width=0.99\linewidth]{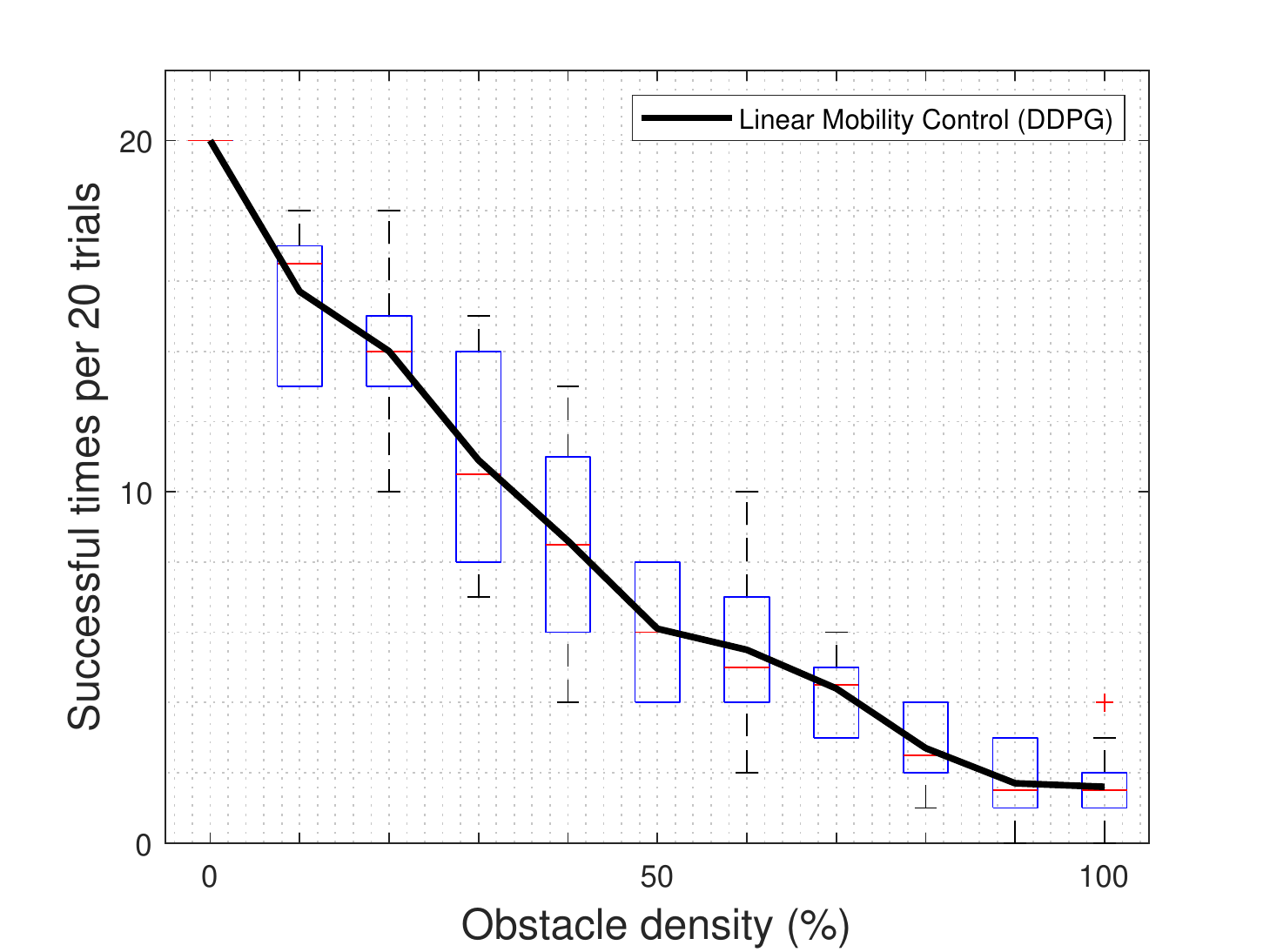}\captionsetup{justification=centering}
     \subcaption{\textbf{Comparison}}
     \end{multicols}
 \caption{Number of successful attacks per 20 trials according to obstacle density.}%
\label{fig:NLNR_vs.LNR}
\end{figure*}

Fig.~\ref{fig:NLNR_vs.LNR} shows the number of successful attacks according to the density of obstacles. In the cyber-physical visualization environment, each experiment was performed for 10 rounds, 20 times in one environment, and the number of times that the target was successfully reached/destroyed was measured in each round. If there are no obstacles, the drone/agent successfully reaches the target/tank in all trials, however the accuracy decreases as the density of the obstacles increases.
Notice that the comparison between Fig.~\ref{fig:NLNR_vs.LNR}(a) and Fig.~\ref{fig:NLNR_vs.LNR}(b) is presented in Appendix~\ref{sec:appendix}. Moreover, this results are numerically presented in Table~\ref{tab:nonlinear} and Table~\ref{tab:linear}. When the obstacle density becomes 50\%, our comparing LMC algorithm shows about 48\% less performance than our proposed situation-aware autonomous nonlinear drone mobility control algorithm. In addition, even with the maximum number of obstacles deployed, our proposed algorithm succeeded in attacking at least four times. This matches the maximum number of successful attacks of conventional DDPG algorithms in the same environment and shows a maximum attack success rate of 2.5 times better than the rate of conventional DDPG algorithm. In particular, in an environment with high obstacle density, our comparing LMC algorithm is almost impossible to use, whereas our proposed situation-aware autonomous nonlinear drone mobility control algorithm can be operated in the same environment. In Table~\ref{tab:comparison}, the performance difference between the two algorithms is compared and summarized. Overall, the higher density of obstacles introduces better performance improvements. When the obstacle density exceeds 70\%, the performance improvement becomes at least 100\%, as presented in Table~\ref{tab:comparison}.


\begin{table*}[h!]
\caption{The number of successful attacks per 20 trials based on obstacle density through our proposed situation-aware autonomous nonlinear drone mobility control.}
\label{tab:nonlinear} 
\resizebox{\textwidth}{!}{%
\begin{tabular}{l r r r r r r r r r r r}
\toprule[1pt] 
\multicolumn{1}{l}{\textbf{Nonlinear Mobility Control (Proposed)}} &
  \textbf{0\%} & \textbf{10\%} & \textbf{20\%} & \textbf{30\%} & \textbf{40\%} & \textbf{50\%} & \textbf{60\%} & \textbf{70\%} & \textbf{80\%} & \textbf{90\%} & \textbf{100\%} \\ \midrule \
Round 1  & 20/20 & 18/20 & 16/20 & 14/20 & 14/20 & 12/20 & 10/20 & 13/20 & 10/20 & 8/20  & 7/20  \\ \ 
Round 2  & 20/20 & 15/20 & 17/20 & 17/20 & 15/20 & 10/20 & 7/20  & 9/20  & 8/20  & 7/20  & 7/20  \\ \
Round 3  & 20/20 & 18/20 & 13/20 & 12/20 & 17/20 & 13/20 & 8/20  & 9/20  & 11/20 & 5/20  & 5/20  \\ \
Round 4  & 20/20 & 16/20 & 18/20 & 17/20 & 16/20 & 11/20 & 9/20  & 6/20  & 7/20  & 8/20  & 7/20  \\ \
Round 5  & 20/20 & 17/20 & 17/20 & 14/20 & 16/20 & 12/20 & 14/20 & 11/20 & 5/20  & 10/20 & 10/20 \\ \
Round 6  & 20/20 & 18/20 & 17/20 & 18/20 & 15/20 & 12/20 & 6/20  & 6/20  & 11/20 & 9/20  & 8/20  \\ \
Round 7  & 20/20 & 18/20 & 17/20 & 17/20 & 16/20 & 12/20 & 14/20 & 13/20 & 6/20  & 11/20 & 10/20 \\ \
Round 8  & 20/20 & 15/20 & 19/20 & 17/20 & 17/20 & 14/20 & 8/20  & 9/20  & 13/20 & 5/20  & 4/20  \\ \
Round 9  & 20/20 & 19/20 & 15/20 & 16/20 & 16/20 & 13/20 & 12/20 & 7/20  & 7/20  & 5/20  & 5/20  \\ \
Round 10 & 20/20 & 18/20 & 17/20 & 16/20 & 10/20 & 10/20 & 11/20 & 8/20  & 8/20  & 9/20  & 10/20 \\ \midrule \
\textbf{Maximum} & \textbf{20} & \textbf{19} & \textbf{19} & \textbf{18} & \textbf{17} & \textbf{14} & \textbf{14} & \textbf{13} & \textbf{13} & \textbf{11} & \textbf{10} \\ \
\textbf{Average} & \textbf{20} & \textbf{17.2} & \textbf{16.6} & \textbf{15.8} & \textbf{15.2} & \textbf{11.9} & \textbf{9.9} & \textbf{9.1} & \textbf{8.6} & \textbf{7.7} & \textbf{7.3} \\ \
\textbf{Median} & \textbf{20} & \textbf{18} & \textbf{17} & \textbf{16.5} & \textbf{16} & \textbf{12} & \textbf{9.5} & \textbf{9} & \textbf{8} & \textbf{8} & \textbf{7} \\ \
\textbf{Minimum} & \textbf{20} & \textbf{15} & \textbf{13} & \textbf{12} & \textbf{10} & \textbf{10} & \textbf{6} & \textbf{6} & \textbf{5} & \textbf{5} & \textbf{4} \\ 
\bottomrule[1pt]
\end{tabular}%
}
\end{table*}

\begin{table*}[h!]
\caption{The number of successful attacks per 20 trials based on obstacle density through our comparing linear mobility control (LMC) algorithm.}
\label{tab:linear} 
\resizebox{\textwidth}{!}{%
\begin{tabular}{l r r r r r r r r r r r}
\toprule[1pt] 
\multicolumn{1}{l}{\textbf{Linear Mobility Control (DDPG)}} &
  \textbf{0\%} & \textbf{10\%} & \textbf{20\%} & \textbf{30\%} & \textbf{40\%} & \textbf{50\%} & \textbf{60\%} & \textbf{70\%} & \textbf{80\%} & \textbf{90\%} & \textbf{100\%} \\ \midrule \
Round 1  & 20/20 & 13/20 & 17/20 & 15/20 & 9/20  & 4/20 & 5/20  & 6/20 & 4/20 & 2/20 & 2/20 \\ \
Round 2  & 20/20 & 17/20 & 13/20 & 13/20 & 10/20 & 8/20 & 4/20  & 4/20 & 2/20 & 1/20 & 1/20 \\ \
Round 3  & 20/20 & 17/20 & 15/20 & 14/20 & 6/20  & 6/20 & 4/20  & 5/20 & 2/20 & 3/20 & 1/20 \\ \
Round 4  & 20/20 & 13/20 & 14/20 & 9/20  & 11/20 & 8/20 & 2/20  & 6/20 & 3/20 & 2/20 & 2/20 \\ \
Round 5  & 20/20 & 18/20 & 18/20 & 10/20 & 12/20 & 6/20 & 9/20  & 3/20 & 1/20 & 3/20 & 1/20 \\ \
Round 6  & 20/20 & 15/20 & 13/20 & 14/20 & 7/20  & 8/20 & 10/20 & 4/20 & 4/20 & 3/20 & 0/20 \\ \
Round 7  & 20/20 & 13/20 & 11/20 & 8/20  & 8/20  & 4/20 & 5/20  & 5/20 & 3/20 & 1/20 & 2/20 \\ \
Round 8  & 20/20 & 17/20 & 15/20 & 11/20 & 6/20  & 4/20 & 5/20  & 5/20 & 2/20 & 1/20 & 3/20 \\ \
Round 9  & 20/20 & 18/20 & 10/20 & 7/20  & 4/20  & 7/20 & 4/20  & 3/20 & 2/20 & 0/20 & 0/20 \\ \
Round 10 & 20/20 & 16/20 & 14/20 & 8/20  & 13/20 & 6/20 & 7/20  & 3/20 & 4/20 & 1/20 & 4/20 \\ 
\midrule \
\textbf{Maximum} &  \textbf{20} &  \textbf{18} &  \textbf{18} &  \textbf{15} &  \textbf{13} &  \textbf{8} &  \textbf{10} &  \textbf{6} &  \textbf{4} &  \textbf{3} &  \textbf{4} \\ \
\textbf{Average} & \textbf{20} & \textbf{15.7} & \textbf{14} & \textbf{10.9} & \textbf{8.6} & \textbf{6.1} & \textbf{5.5} & \textbf{4.4} & \textbf{2.7} & \textbf{1.7} & \textbf{1.6} \\ \
\textbf{Median} &  \textbf{20} &  \textbf{16.5} &  \textbf{14} &  \textbf{10.5} &  \textbf{8.5} &  \textbf{6} &  \textbf{5} &  \textbf{4.5} &  \textbf{2.5} &  \textbf{1.5} &  \textbf{1.5} \\ \
\textbf{Minimum} &  \textbf{20} &  \textbf{13} &  \textbf{10} &  \textbf{7} &  \textbf{4} &  \textbf{4} &  \textbf{2} &  \textbf{3} &  \textbf{1} &  \textbf{0} &  \textbf{0} \\
  \bottomrule[1pt]
\end{tabular}%
}
\end{table*}

\begin{table*}[h!]
\caption{Comparison of our proposed nonlinear control and the comparing linear control algorithm.}
\label{tab:comparison} 
\resizebox{\textwidth}{!}{%
\begin{tabular}{l r r r r r r r r r r r}
\toprule[1pt] 
\multicolumn{1}{l}{\textbf{Performance Improvement(\%)}} &
  \textbf{0\%} & \textbf{10\%} & \textbf{20\%} & \textbf{30\%} & \textbf{40\%} & \textbf{50\%} & \textbf{60\%} & \textbf{70\%} & \textbf{80\%} & \textbf{90\%} & \textbf{100\%} \\ 
\midrule \
\textbf{Maximum} &  {0} &     {5.5} &    {5.5} &    {20} &    {30.77} &    {75} &    {40} &    {116.67} &    {225} &    {266.67} &    {150}  \\ \
\textbf{Average} &  {0} &    {9.55} &   {18.57} &   {44.95} &   {76.74} &   {95.08} &   {80} &   {106.82} &   {218.52} &   {352.94} &   {356.25}  \\ \
\textbf{Median} &  {0} &     {9.09} &    {21.43} &    {57.14} &    {88.24} &    {100} &    {90} &    {100} &    {220} &    {433.33} &    {366.67} \\ \
\textbf{Minimum} &  {0} &    {15.38} &    {30} &    {71.43} &    {150} &    {150} &    {200} &    {100} &    {400} &    {-} &    {-} \\
  \bottomrule[1pt]
\end{tabular}%
}
\end{table*}

\section{Concluding Remarks and Future Work}\label{sec:conclusion}
This paper proposes a novel situation-aware autonomous nonlinear drone mobility control DRL-based algorithm in cyber-physical loitering munition applications. 
On the battlefield, the design and implementation of a DRL-based algorithm are not straightforward because real-world data gathering is not easy at all. Therefore, the cyber-physical virtual environment is constructed with Unity in this paper. Based on that, a DRL-based automated drone mobility control algorithm can be designed, evaluated, and visualized. 
In real world battlefield scenarios, many obstacles exist which is harmful to linear trajectory control. Therefore, this paper proposes a novel nonlinear drone mobility control using situation-aware components (e.g., Raycast function in Unity for the virtual environment). Based on the sensed situation-aware information, the drone can adjust its trajectory during flight. Therefore, this approach can be definitely beneficial for avoiding obstacles on the battlefield. Our visualization-based performance evaluation shows that the proposed algorithm is superior to the other conventional linear mobility control algorithms. 

As future research directions, the proposed algorithm can be implemented on top of embedded drone platforms where the platforms are equipped with multiple sensors for situation-awareness functionalities.}

\bibliographystyle{IEEEtran}
\bibliography{ref_aimlab, ref_cps}

\appendix
\section{Performance Comparison}\label{sec:appendix}
\begin{figure}[t!]
    \centering
    \includegraphics[width =0.99\linewidth]{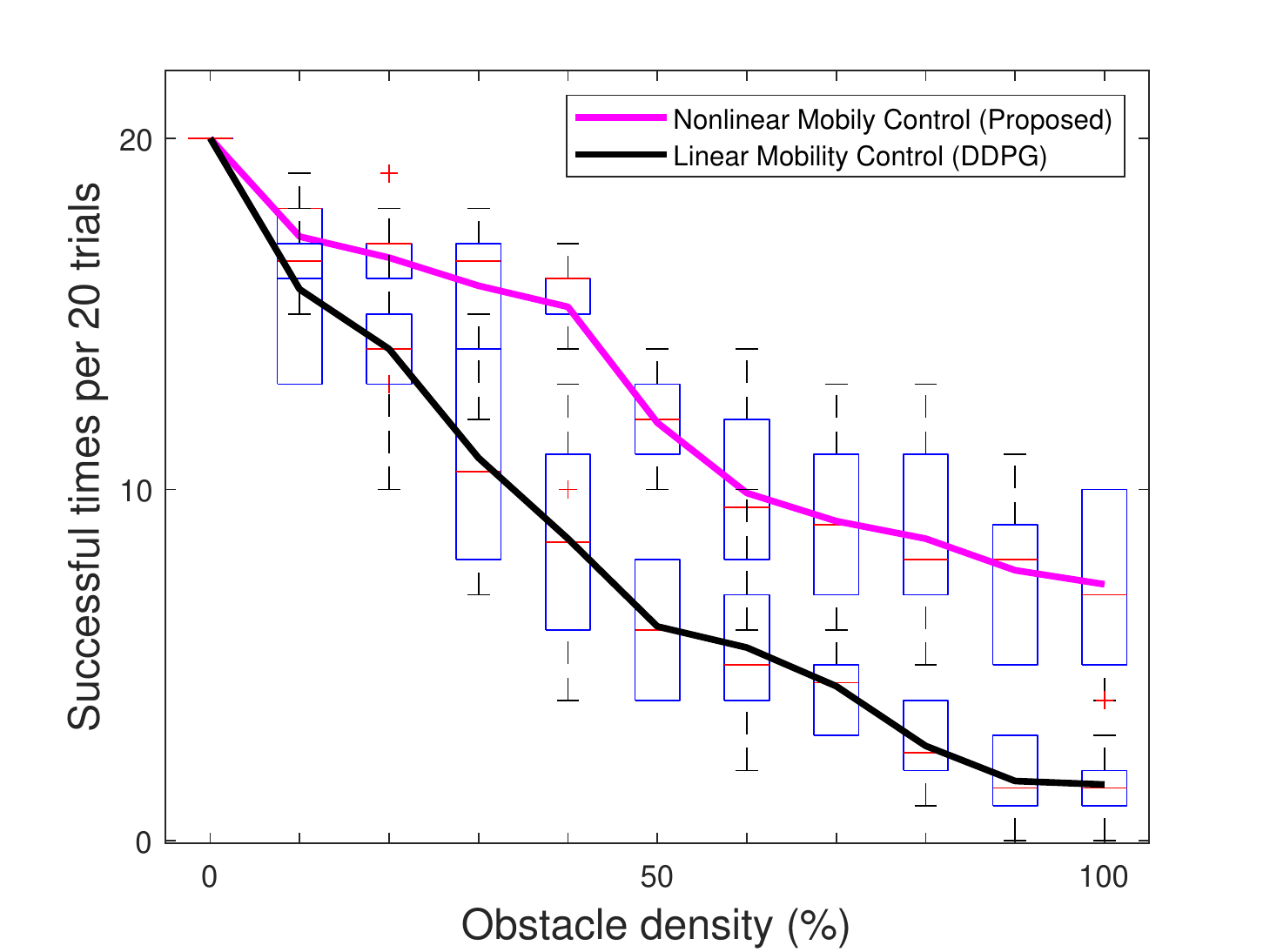}
    \caption{Number of successful attacks per 20 trials according to obstacle density}
    	\label{fig:NLNR_vs.LNR_total}
\end{figure}

In Fig.~\ref{fig:NLNR_vs.LNR_total}, we can compare the performance of the two algorithms, i.e., our proposed nonlinear drone mobility control and the comparing linear drone mobility control at a glance. As the obstacle density increases, performance degradation occurs in both algorithms. Therefore, we can confirm that our proposed nonlinear mobility control has much less performance degradation. Furthermore, we can also confirm that our proposed algorithm becomes getting better when the density of obstacles increases, i.e., the performance gap between the proposed nonlinear mobility control algorithm and the comparing linear mobility control becomes larger when the density of obstacles increases. 

\begin{IEEEbiography}[{\includegraphics[width=1in,height=1.25in,clip]{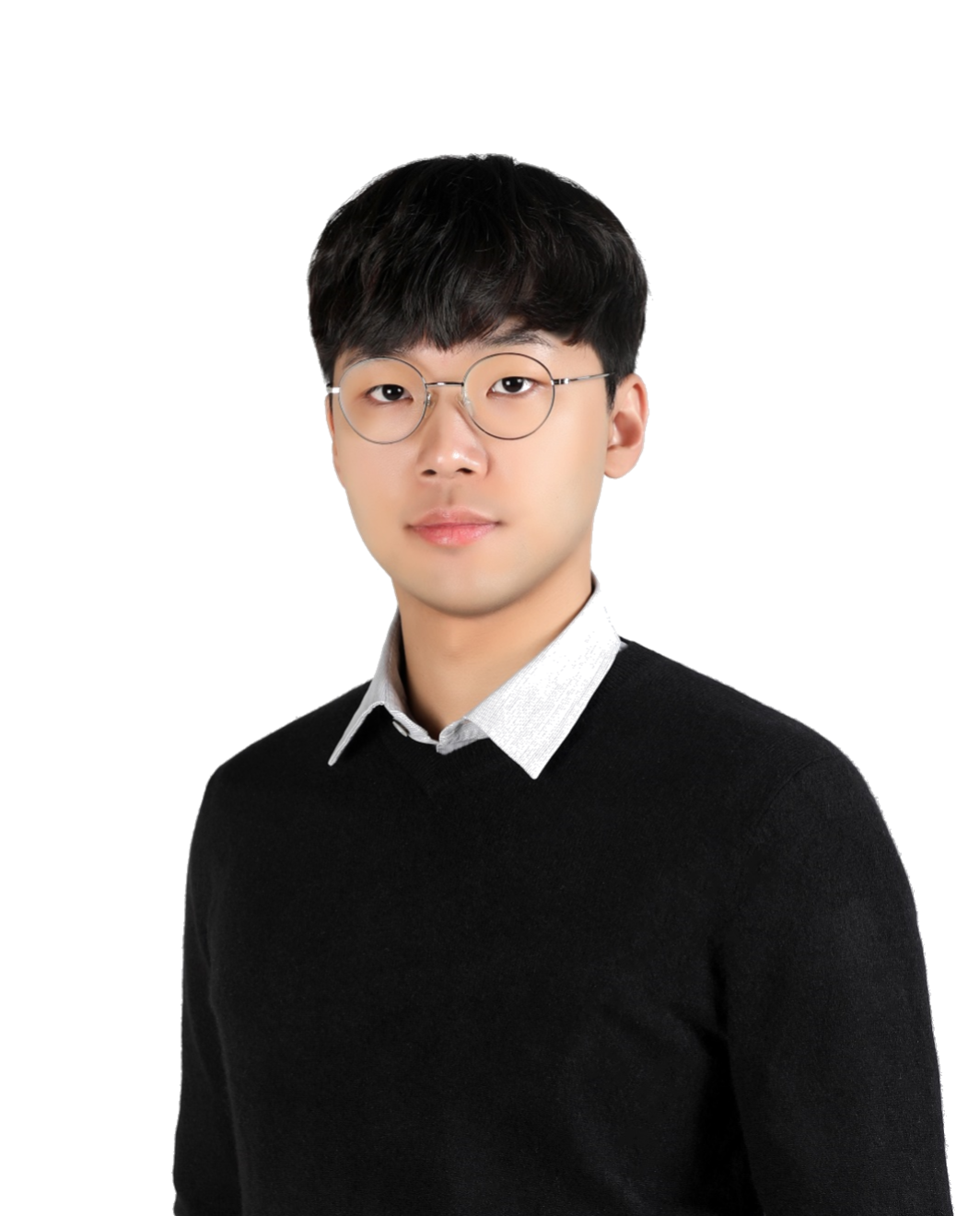}}]{Hyunsoo Lee} is currently pursuing the Ph.D. degree in electrical and computer engineering at Korea University, Seoul, Republic of Korea. He received the B.E. degree in electronic engineering from Soongsil University, Seoul, Republic of Korea, in 2021. His research focuses include deep learning algorithms and their applications to mobility and networking. 

He was a recipient of the IEEE Vehicular Technology Society (VTS) Seoul Chapter Award in 2022.
\end{IEEEbiography}

\begin{IEEEbiography}[{\includegraphics[width=1in,height=1.25in,clip]{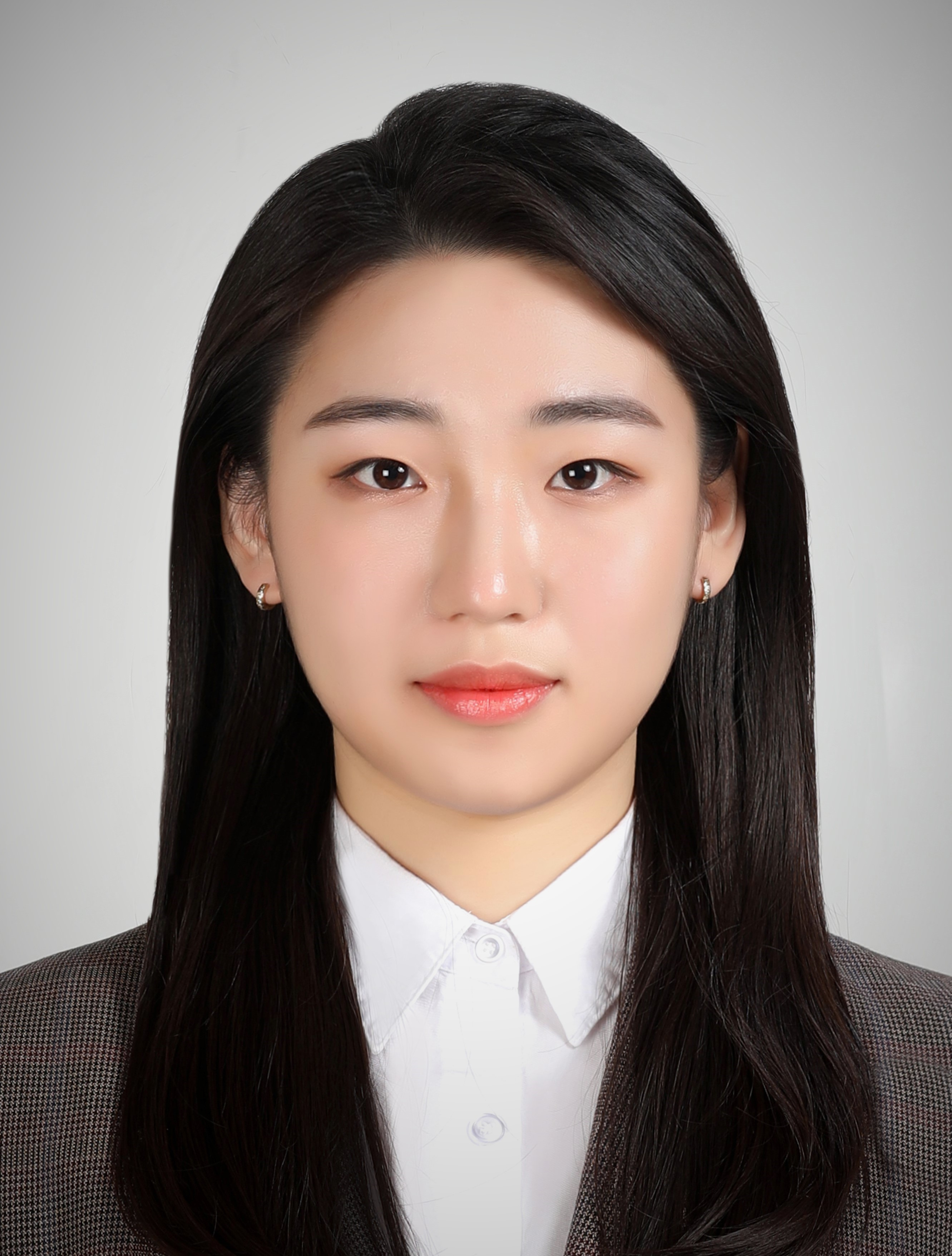}}]{Soohyun Park} 
is currently pursuing the Ph.D. degree in electrical and computer engineering at Korea University, Seoul, Republic of Korea. She received the B.S. degree in computer science and engineering from Chung-Ang University, Seoul, Republic of Korea, in 2019. Her research focuses include deep learning algorithms and their applications to autonomous mobility and connected vehicles. 

She was a recipient of the IEEE Vehicular Technology Society (VTS) Seoul Chapter Award in 2019.
\end{IEEEbiography}

\begin{IEEEbiography}[{\includegraphics[width=1in,height=1.25in,clip]{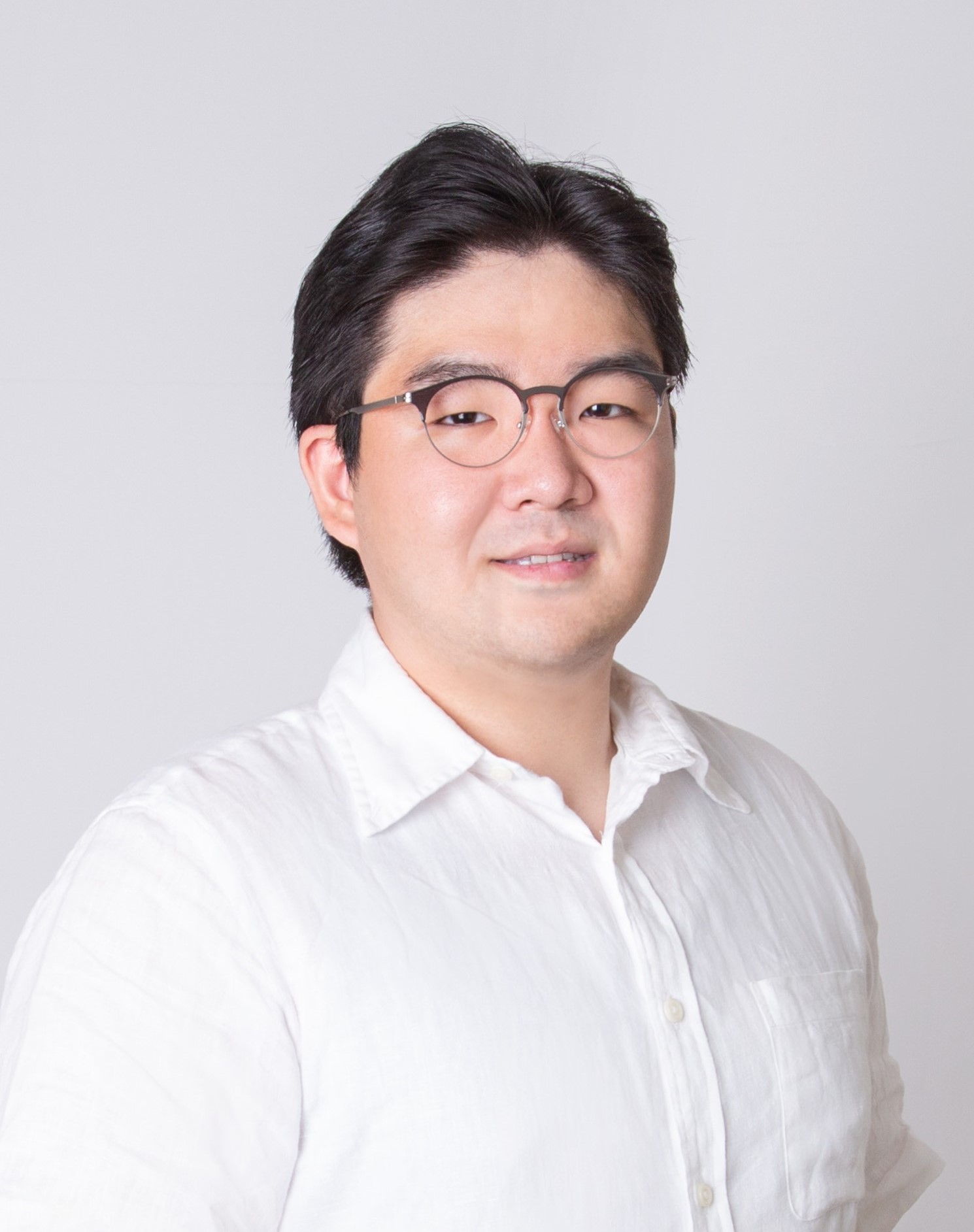}}]{Won Joon Yun} 
 is currently a Ph.D. student in electrical and computer engineering at Korea University, Seoul, Republic of Korea, since March 2021, where he received his B.S. in electrical engineering. He was a visiting researcher at Cipherome Inc., San Jose, CA, USA (Summer 2022),; and also a visiting researcher at the University of Southern California (USC), Los Angeles, CA, USA (Spring 2023) for a joint project with Prof. Andreas F. Molisch at the Ming Hsieh Department of Electrical and Computer Engineering, USC Viterbi School of Engineering. 

 He was a recipient of IEEE ICOIN Best Paper Award (2021).
\end{IEEEbiography}


\begin{IEEEbiography}[{\includegraphics[width=1in,height=1.25in,clip,keepaspectratio]{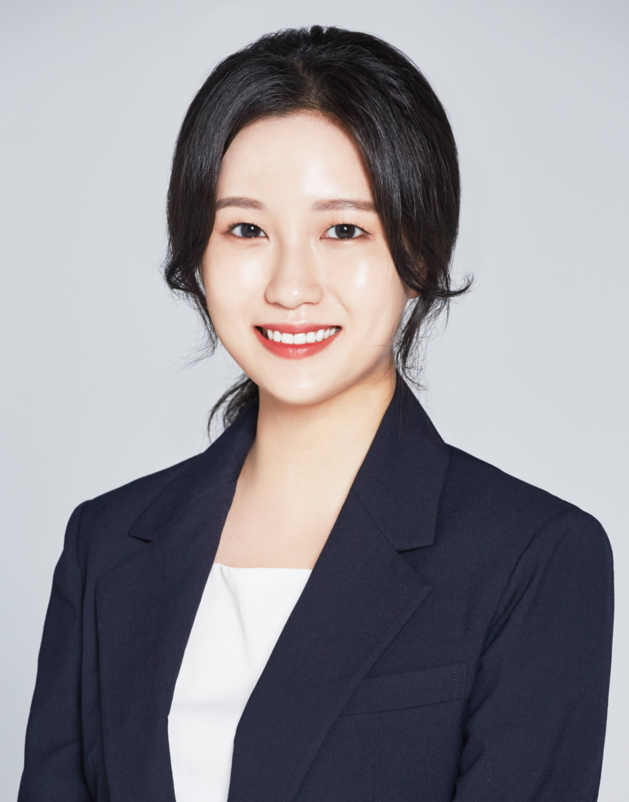}}]{Soyi Jung} has been an assistant professor at the Department of Electrical of Computer Engineering, Ajou University, Suwon, Republic of Korea, since September 2022. Before joining Ajou University, she was an assistant professor at Hallym University, Chuncheon, Republic of  Korea, from 2021 to 2022; a visiting scholar at Donald Bren School of Information and Computer Sciences, University of California, Irvine, CA, USA, from 2021 to 2022; a research professor at Korea University, Seoul, Republic of Korea, in 2021; and a researcher at Korea Testing and Research (KTR) Institute, Gwacheon, Republic of Korea, from 2015 to 2016. She received her B.S., M.S., and Ph.D. degrees in electrical and computer engineering from Ajou University, Suwon, Republic of Korea, in 2013, 2015, and 2021, respectively. Her current research interests include network optimization for autonomous vehicles communications, distributed system analysis, big-data processing platforms, and probabilistic access analysis. She was a recipient of Best Paper Award by KICS (2015), Young Women Researcher Award by WISET and KICS (2015), Bronze Paper Award from IEEE Seoul Section Student Paper Contest (2018), ICT Paper Contest Award by Electronic Times (2019), and IEEE ICOIN Best Paper Award (2021).
\end{IEEEbiography}

\begin{IEEEbiography}[{\includegraphics[width=1in,height=1.25in,clip,keepaspectratio]{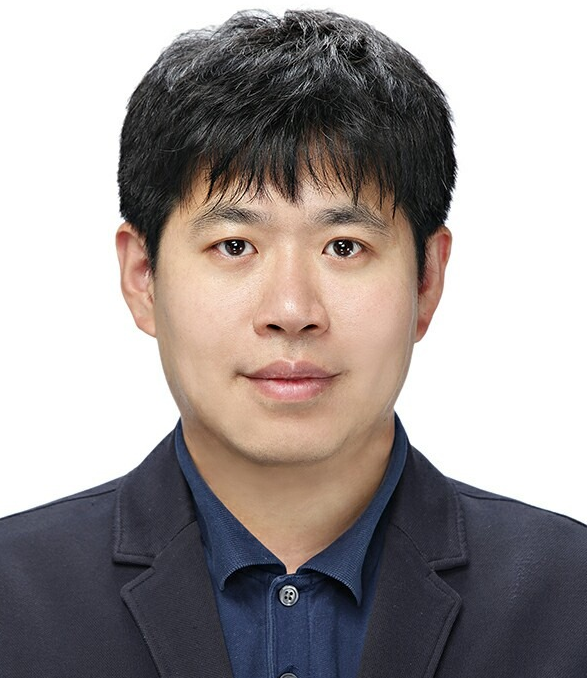}}]{Joongheon Kim}
(M'06--SM'18) has been with Korea University, Seoul, Korea, since 2019, where he is currently an associate professor at the School of Electrical Engineering and also an adjunct professor at the Department of Communications Engineering (established/sponsored by Samsung Electronics) and the Department of Semiconductor Engineering (established/sponsored by SK Hynix). He received the B.S. and M.S. degrees in computer science and engineering from Korea University, Seoul, Korea, in 2004 and 2006; and the Ph.D. degree in computer science from the University of Southern California (USC), Los Angeles, CA, USA, in 2014. Before joining Korea University, he was a research engineer with LG Electronics (Seoul, Korea, 2006--2009), a systems engineer with Intel Corporation (Santa Clara, CA, USA, 2013--2016), and an assistant professor of computer science and engineering with Chung-Ang University (Seoul, Korea, 2016--2019). 

He serves as an editor for \textsc{IEEE Transactions on Vehicular Technology}, \textsc{IEEE Transactions on Machine Learning in Communications and Networking}, and \textsc{IEEE Communications Standards Magazine}. He is also a distinguished lecturer for \textit{IEEE Communications Society (ComSoc)} and \textit{IEEE Systems Council}.

He was a recipient of Annenberg Graduate Fellowship with his Ph.D. admission from USC (2009), Intel Corporation Next Generation and Standards (NGS) Division Recognition Award (2015), \textsc{IEEE Systems Journal} Best Paper Award (2020), IEEE ComSoc Multimedia Communications Technical Committee (MMTC) Outstanding Young Researcher Award (2020), IEEE ComSoc MMTC Best Journal Paper Award (2021), and Best Special Issue Guest Editor Award by \textit{ICT Express (Elsevier)} (2022). He also received several awards from IEEE conferences including IEEE ICOIN Best Paper Award (2021), IEEE Vehicular Technology Society (VTS) Seoul Chapter Awards (2019, 2021, and 2022), and IEEE ICTC Best Paper Award (2022). 
\end{IEEEbiography}
\end{document}